\documentclass[aps,prd,twocolumn,showpacs,superscriptaddress,nofootinbib,preprintnumbers]{revtex4-2}
\usepackage{amsmath}

\usepackage{tikz}
\usepackage{tikz-feynman}
\tikzfeynmanset{
  compat=1.1.0
}
\tikzset{every path/.append style={thick}}
\usepackage{amsfonts}
\usepackage{amssymb}
\usepackage{enumerate}
\usepackage{color}
\usepackage{graphicx}
\usepackage{bm}
\usepackage[colorlinks=true,citecolor=blue,urlcolor=blue]{hyperref}
\usepackage{afterpage}
\usepackage{multirow}
\usepackage{appendix}
\usepackage[table]{xcolor}
\usepackage{feynmp-auto}
\definecolor{tabblue}{RGB}{31, 119, 180}
\definecolor{darkblue}{RGB}{0, 0, 120}
\definecolor{tabred}{RGB}{214, 39, 40}
\definecolor{tabgreen}{RGB}{44, 160, 44}
\definecolor{tabgray}{RGB}{100, 100, 100}
\usepackage{float}
\usepackage{placeins}

\begin{document}
\preprint{UTWI-02-2025}

\title{CMB Constraints on Loop-Induced Decays of Leptophilic Dark Matter}

\author{Gabriele Montefalcone}\email{montefalcone@utexas.edu}
\affiliation{Texas Center for Cosmology and Astroparticle Physics, Weinberg Institute, Department of Physics, The University of Texas at Austin, Austin, TX 78712, USA}

\author{Gilly Elor}
\affiliation{Texas Center for Cosmology and Astroparticle Physics, Weinberg Institute, Department of Physics, The University of Texas at Austin, Austin, TX 78712, USA}

\author{Kimberly K. Boddy}
\affiliation{Texas Center for Cosmology and Astroparticle Physics, Weinberg Institute, Department of Physics, The University of Texas at Austin, Austin, TX 78712, USA}

\author{Nicola Bellomo}
\affiliation{Texas Center for Cosmology and Astroparticle Physics, Weinberg Institute, Department of Physics, The University of Texas at Austin, Austin, TX 78712, USA}
\affiliation{Dipartimento di Fisica e Astronomia G. Galilei, Universit\`a degli Studi di Padova, Via Marzolo 8, I-35131 Padova, Italy}
\affiliation{INFN, Sezione di Padova, Via Marzolo 8, I-35131, Padova, Italy}
\affiliation{INAF - Osservatorio Astronomico di Padova, Vicolo dell'Osservatorio 5, I-35122 Padova, Italy}

\begin{abstract}

Leptophilic sub-MeV spin-zero dark matter (DM) decays into photons via one-loop processes, a scenario that has been in part overlooked in current literature.
In this work, we provide updated and comprehensive upper limits on scalar, pseudo-scalar, and axion-like DM-electron couplings based on the latest \texttt{NPIPE} cosmic microwave background data from \textit{Planck}. Our bounds on the couplings are not only competitive with astrophysical and terrestrial experiments, but outperform them in certain regions of parameter space.
Notably, we present the most stringent limits to date on scalar DM with masses around a few keV and pseudo-scalar DM with masses between 100 eV and a few keV.
Additionally, we explore, for the first time, the impact of implementing a cosmology-consistent treatment of energy deposition into the cosmic medium.
\end{abstract}

\maketitle

\section{Introduction}
\label{sec:intro}

Uncovering the nature and origin of dark matter (DM) is one of the most outstanding problems in particle physics today.
Terrestrial experiments (e.g., particle colliders and direct detection experiments) have yet to discover the elusive DM particle(s) or shed light on any non-gravitational interactions it may have.
Non-detections place stringent bounds on the DM mass and interactions with Standard Model (SM) particles, thereby restricting the parameter space of weakly interacting massive particles and motivating the study of new DM candidates beyond this paradigm.

Scalar and pseudoscalar DM models, including axions and axion-like particles (ALPs), appear in many well motivated extensions of the SM~\cite{PhysRevLett.40.223, Adams:2022pbo,Choi:2020rgn,Raffelt:2006rj}.
In these scenarios, the full DM abundance can be generated through non-thermal mechanisms, such as field misalignment or topological defect decays~\cite{Batell:2022qvr, PRESKILL1983127, DINE1983137, DAVIS1986225, Gorghetto:2018myk, Gorghetto:2020qws, Lyth:1991ub, Co:2019jts, Co:2018mho, Co:2020dya, Cembranos:2019qlm}, permitting the production of sub-MeV DM without violating bounds from big bang nucleosynthesis (BBN) or cosmic microwave background (CMB) anisotropies~\cite{Boehm:2002yz,Serpico:2004nm,Nollett:2013pwa,Steigman:2014pfa,Nollett:2014lwa,Escudero:2018mvt,Sabti:2019mhn,Giovanetti:2021izc,An:2022sva,An:2024nsw}.

In this work, we consider the cosmological implications of spin-zero DM candidates with sub-MeV masses that couple trilinearly at tree level to SM electrons.
This class of DM models has been considered previously, especially in the context of direct and indirect detection~\cite{Mitridate:2021ctr, Knapen:2017xzo,Essig:2013goa}, but the detailed impact on cosmological observables was largely overlooked.
We note that, contrary to naive expectations, the most frequent DM-electron process in a cosmological setup is not the 2-to-2 tree-level DM-electron scattering, but the (electron) loop-induced DM decay to two photons.
Although the decay rate needs to be sufficiently low to render DM cosmologically stable, the process of DM decay into SM particles injects energy into the cosmic medium, changing the cosmological thermal and ionization history.
CMB anisotropies, for example, are very sensitive to such changes and are thus a powerful probe of DM decay.

The impact of DM decays on cosmology has been extensively studied~\cite{Adams:1998nr, Chen:2003gz, 2012PhRvD..85d3522F,Slatyer:2012yq, Poulin:2016nat,Slatyer:2016qyl,Liu:2016cnk, Poulin:2016anj, Cang:2020exa, Liu:2023nct, Capozzi:2023xie, 2024arXiv240813305X}. Recent work~\cite{Capozzi:2023xie,Liu:2023nct,2024arXiv240813305X} placed constraints on the decay of sub-MeV DM using the \texttt{DarkHistory} code~\cite{Liu:2019bbm, Liu:2023fgu} to calculate the thermal history with exotic energy injection and interfacing it with \texttt{ExoCLASS}~\cite{Stocker:2018avm}, an extension of the Boltzmann solver \texttt{CLASS}~\cite{Blas:2011rf}.

We improve upon the current theoretical landscape by considering the impact of decaying sub-MeV spin-zero leptophilic DM on CMB anisotropies.
Moreover, we build on the analysis of Ref.~\cite{2024arXiv240813305X}, extending it in two ways.
First, we investigate, for the first time, the impact of a cosmologically consistent treatment of the energy deposition process.
Since \texttt{DarkHistory} has the standard cosmological parameters internally hard coded, it computes the energy deposition for a fixed standard cosmology.
We modify \texttt{DarkHistory} to allow our Markov chain Monte Carlo (MCMC) to explore the full cosmological parameter space with a self-consistent thermal history.
We use the \textit{Planck} 2018 PR3 likelihood to set constraints on DM-electron interactions in the sub-MeV regime for both the fixed and cosmologically consistent treatment in \texttt{DarkHistory}. Second, we test the impact of the latest \textit{Planck} \texttt{NPIPE} data release~\cite{Planck:2020olo}, leveraging its improved precision over PR3 in the temperature and polarization spectra to refine constraints on DM-electron interactions.

Through our comprehensive investigation, we obtain robust and reliable constraints on the decay of sub-MeV scalar, pseudo-scalar, and ALP DM from CMB anisotropies.
We find that CMB anisotropies provide competitive constraints on DM-electron couplings for scalar and pseudo-scalar DM.
We improve upon previous astrophysical and terrestrial bounds for scalar DM with mass around a few keV and for pseudo-scalar DM in the mass range of 100~eV to $\sim 4$~keV.
For ALP DM, while we do not improve upon existing constraints, our bounds complement those obtained from direct and indirect detection.
Although we focus on DM-electron coupling, we also apply our results to well-motivated DM models with couplings to muons and tau-leptons.

This paper is organized as follows. 
In Sec.~\ref{sec:models}, we introduce the three DM models under consideration---scalar DM, pure pseudo-scalar DM, and ALPs---and provide their 1-loop decay rates. 
In Sec.~\ref{sec:energy_injection}, we briefly describe the methodology for calculating the effects of exotic energy injection on the CMB anisotropies, focusing on our use of \texttt{DarkHistory}, and outline our analysis framework in Sec.~\ref{sec:analysis}.
We present our CMB constraints on DM-electron couplings and apply our results to other flavor couplings in Sec.~\ref{sec:results}.
We compare our bounds with existing limits from astrophysical sources and terrestrial experiments in Sec.~\ref{sec:comparison}.
Finally, we conclude and highlight future directions in Sec.~\ref{sec:conclusions}.

\section{Models}
\label{sec:models}

We consider scalar, pseudo-scalar, and ALP particles that couple to electrons and constitute all of the DM in the universe.
In all cases of interest, we restrict the DM mass to be $m_\mathrm{DM} < 2 m_e$, where $m_e$ is the mass of the electron, so that the tree-level decay of DM into electrons is kinematically forbidden.

We remain agnostic about the specific DM production mechanism. Consequently, we treat the relic abundance as having been set earlier by an independent non-thermal process (e.g., field misalignment), rather than by the electron coupling explored here.
Instead, we focus on the low-energy phenomenology of DM at later times, around the CMB era, allowing our study to be effectively independent of the UV completion of the DM models. However, we do assume DM is produced non-thermally, and DM must remain thermally decoupled from the SM before and around the time of BBN to avoid constraints on light element abundances from BBN and CMB anisotropies~\cite{Boehm:2002yz,Serpico:2004nm,Nollett:2013pwa,Steigman:2014pfa,Nollett:2014lwa,Escudero:2018mvt,Sabti:2019mhn,Giovanetti:2021izc,An:2022sva,An:2024nsw}.
For each model of interest, we verify that our upper bounds on the strength of DM-electron couplings are sufficiently small such that DM remains non-thermalized with the SM bath at temperatures $T \gtrsim \mathrm{MeV}$.

Additionally, the interactions we consider permit both DM decay and inelastic scattering, in which DM is in the initial but not the final state.
The decay lifetimes of interest are much longer than the age of the universe and thus do not alter the DM density in any appreciable way.
We have verified that inelastic scattering does not impact the DM number density; the scattering rates are smaller than the Hubble expansion rate at temperatures $T \simeq m_\mathrm{DM}$.

\begin{figure}[ht!]
  \centering
  \tikzset{every fermion/.style={very thick}}
  \begin{tikzpicture}
    \begin{feynman}
      \vertex (i)  at (0,0)         {\(\phi,\Phi,a\)};
      \vertex[dot] (v1) at (2,0)     {};
      \vertex[dot] (v2) at (4,1.5)   {};
      \vertex[dot] (v3) at (4,-1.5)  {};
      \vertex (o1) at (6,1.5)       {\(\gamma\)};
      \vertex (o2) at (6,-1.5)      {\(\gamma\)};
      \diagram* {
        (i)  -- [scalar] (v1),
        (v1) -- [fermion] (v2),
        (v2) -- [fermion] (v3),
        (v3) -- [fermion] (v1),
        (v2) -- [photon]  (o1),
        (v3) -- [photon]  (o2),
      };
      \end{feynman}
  \end{tikzpicture}
  \caption{Feynman diagram for the electron-mediated one-loop decay of scalar~($\phi$), pseudo-scalar~($\Phi$), or axion-like particle~($a$) DM into two photons.  This decay channel represents the dominant interaction responsible for observable cosmological signatures analyzed in this work.
  }
  \label{fig:loop_diagram}
\end{figure}

\subsection{Scalar}
The interaction Lagrangian for a light, leptophilic scalar DM particle, $\phi$, coupling to electrons is
\begin{equation}
  \mathcal{L}_{\rm{int}} =  g_{\phi \bar e e} \phi \bar\psi_e \psi_e\,,
  \label{eq:Lint_scalar}
\end{equation}
where~$g_{\phi \bar{e} e}$ is the scalar DM-electron coupling constant.
At leading order, $\phi$ decays into two photons through an electron loop, illustrated in Fig.~\ref{fig:loop_diagram}, with corresponding decay width
\begin{equation}
  \Gamma_{\phi \rightarrow \gamma \gamma} = \frac{g_{\phi \bar e e}^2 \alpha_{\rm{EM}}^2}{288 \pi^3} \frac{m_\phi^3}{m_e^2} + \mathcal{O}\left(\frac{1}{m_e^4}\right),
  \label{eq:Gamma_scalar}
\end{equation}
where~$\alpha_\mathrm{EM}$ is the electromagnetic fine-structure constant, and~$m_\phi$ is the mass of scalar DM.
The 1-loop decay rate dominates over the $\phi e \rightarrow \gamma e$ scattering rate by at least 20 orders of magnitude at redshift~$z \lesssim 10^4$, making it the only contribution of interest for our analysis.
Even though the 1-loop decay rate is suppressed by an additional factor of~$\alpha_{\rm{EM}}$ compared to the tree-level scattering rate, the latter is substantially more suppressed due to the low electron number density.
Thermalization with SM particles via~$\gamma e \rightarrow e \phi$ and~$ee \rightarrow \gamma \phi$ processes is also suppressed for $m_\phi \ll 1~\text{MeV}$ and $g_{\phi \bar e e} \lesssim 5 \times 10^{-10}$~\cite{Knapen:2017xzo}, and our results lie well below this thermalization threshold.


\subsection{Pseudo-scalar Particles}

The interaction Lagrangian for a light pseudo-scalar DM particle, $\Phi$, coupled to electrons is given by
\begin{equation}
  \mathcal{L}_{\rm{int}} = -g_{\Phi \bar e e} \Phi \bar\psi_e i \gamma_5 \psi_e,
  \label{eq:Lint_pseudoscalar}
\end{equation}
where~$g_{\Phi \bar{e} e}$ is the pseudo-scalar DM-electron coupling constant.
As in the scalar case, the dominant decay channel in the sub-MeV mass range is the 1-loop electron-mediated decay into two photons, with a corresponding rate
\begin{equation}
  \Gamma_{\Phi \rightarrow \gamma \gamma} = \frac{g_{\Phi \bar e e}^2 \alpha_{\rm{EM}}^2}{64 \pi^3} \frac{m_\Phi^3}{m_e^2} + \mathcal{O}\left(\frac{1}{m_e^4}\right),
  \label{eq:Gamma_pseudoscalar}
\end{equation}
where~$m_\Phi$ is the mass of pseudo-scalar DM.
In analogy with the scalar model, we estimate thermalization effects due to scattering processes, such as $\gamma e \rightarrow \Phi e$.
Scattering is suppressed with respect to Hubble expansion for couplings $g_{\Phi\bar{e}e} \lesssim 10^{-10}$ for sub-MeV pseudo-scalars.


\subsection{Axion-Like Particles}

Among other possible pseudo-scalar DM candidates, we single out ALPs, $a$, as a particularly interesting class of models.
The ALP-electron interaction is typically described by the derivative coupling
\begin{equation}
  \mathcal{L}_{\rm{int}} = \frac{g_{a \bar e e}}{2 m_e} \partial_\mu a \bar{\psi}_e \gamma^\mu \gamma_5 \psi_e \,.
  \label{eq:Lint_ALP}
\end{equation}
where~$g_{a \bar{e} e}$ is the ALP DM-electron coupling constant.
At tree-level, the ALP interaction is formally equivalent to the pseudo-scalar interaction of Eq.~\eqref{eq:Lint_pseudoscalar}. 
However, at loop-level, the two interactions do not give rise to the same matrix element due to additional anomalous couplings~\cite{Quevillon:2019zrd, Ferreira:2022xlw}. 
As a result, the decay width of an ALP to two photons via an electron loop is~\cite{Ferreira:2022xlw}
\begin{equation}
  \Gamma_{a\rightarrow \gamma \gamma} =\frac{g_{a \bar e e}^2\alpha_{\rm{EM}}^2}{64 \pi^3} \frac{  m_a^7}{ 144 m_e^6 } \,,
  \label{eq:Gamma_alp}
\end{equation}
for ALP masses~$m_a \ll 2 m_e$.
Similar to the scalar case, ALP-electron scattering processes are suppressed at least ten orders of magnitude compared to the 1-loop decay.
Moreover, the values of $g_{a \bar e e}$ needed for thermalization at temperatures below an MeV are significantly higher than any other bounds we consider~\cite{Green:2021hjh}.

\section{Energy injection and deposition} \label{sec:energy_injection}

The impact of decaying DM is typically parametrized in terms of the rate of energy injection per volume $V$ into the primordial plasma:
\begin{equation}
  \left(\frac{d^2E}{dt\, dV}\right)_{\rm{inj}} =  \Gamma_{\rm{DM}\to\gamma\gamma} e^{-\Gamma_{\rm{DM}\to\gamma\gamma}t} \ \rho_{\rm{DM}}(z),
\end{equation}
where~$t$ is cosmic time, $\Gamma_{\rm{DM}\to\gamma\gamma}$ represents the DM decay rate as defined in Eq.~\eqref{eq:Gamma_scalar}, \eqref{eq:Gamma_pseudoscalar} or~\eqref{eq:Gamma_alp}, and $\rho_{\rm{DM}}(z)$ is the DM energy density at redshift $z$.

Photons produced by DM decays propagate across the cosmic medium while cooling and depositing their energy through various channels: hydrogen and helium ionization, Lyman-$\alpha$ excitations, heating of the intergalactic medium, and the production of low-energy continuum photons. 
The fraction of energy deposited into each channel~$c$ is
\begin{equation}
  \left(\frac{d^2E}{dt\, dV}\right)_{\rm{dep}} = f_c(z) \left(\frac{d^2E}{dt\, dV}\right)_{\rm{inj}},
  \label{eq:f_c}
\end{equation}
where~$f_c(z)$ is the energy deposition function, which captures the physics of radiation transfer.

The main cosmological effect of DM decaying into photons is the overall increase of the free-electron fraction across a wide range of redshifts~\cite{2012PhRvD..85d3522F, Slatyer:2016qyl}.
An increase in free electrons increases the optical depth and a broadens the surface of last scattering compared to the standard cold DM (CDM) scenario.
The former causes an overall suppression of the peaks in the CMB power spectra, while the latter enhances diffusion damping at large multipoles.

We compute the energy deposition functions using version 2.0 of \texttt{DarkHistory}.
This updated version of the code incorporates an improved modeling of low-energy particle deposition, which is highly relevant for the sub-keV energy injections we consider in this work, and a multi-level hydrogen atom approximation in order to track the occupation of multiple excited states.
Following Refs.~\cite{Liu:2023nct, Capozzi:2023xie, 2024arXiv240813305X}, we provide~$f_c(z)$ as an input to the~\texttt{injection} module of~\texttt{CLASS},\footnote{The \texttt{injection} module in \texttt{CLASS} is based on \texttt{ExoCLASS}. With $f_c(z)$ provided by \texttt{DarkHistory}, the \texttt{injection} module and \texttt{ExoCLASS} handle energy deposition identically.} which calculates the evolution of the ionization fraction and baryon temperature under the effects of exotic energy injection.

\begin{figure}[t]
    \centering
    \includegraphics[width=\linewidth]{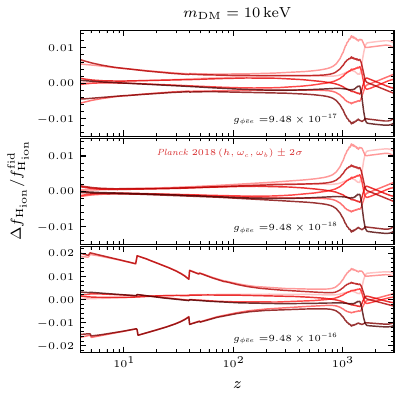}
    \caption{Residuals in the hydrogen ionization channel of the energy deposition function,
$f_{\mathrm{H\,ion}}(z)$, for $m_{\mathrm{DM}}=10\,\mathrm{keV}$ and a given cosmology 
relative to the \textit{Planck}~2018 best-fit baseline. The three panels ({\it top} to {\it bottom})
correspond to scalar DM-electron couplings $g_{\phi \bar{e}e}$ at the 95\%~C.L.\ 
exclusion limit from Ref.~\cite{2024arXiv240813305X} ({\it top}), an order of magnitude 
larger ({\it middle}), and an order of magnitude smaller ({\it bottom}). In each panel, the eight 
curves, shown in various shades of \textcolor{tabred}{red}, represent different combinations of $(h,\,\omega_b,\,\omega_c)$ taken at their 
$\pm 95\%$~C.L.\ \textit{Planck}~2018 values.}
    \label{fig:fs_cosmo_mDM10keV}
\end{figure}

\texttt{DarkHistory} treats the energy deposition functions $f_c(z)$ as independent of cosmology by pre-computing them under a fixed fiducial model. While this approach provides a well-motivated approximation, it is important to note that the $f_c(z)$ are in fact mildly sensitive to the underlying cosmology, primarily through changes in the Hubble constant, $h$, and the physical baryon and DM density, $\omega_b$ and $\omega_c$. Variations in these quantities can alter the redshift evolution of energy deposition by modifying both the expansion history and the abundance of free electrons. 

In Fig.~\ref{fig:fs_cosmo_mDM10keV}, we show how these cosmological parameter variations affect the hydrogen ionization channel, which is the most relevant in setting CMB constraints on decaying DM, in the energy deposition function. We plot residuals in $f_{\mathrm{H_{ion}}}(z)$ for eight combinations of $(h,\,\omega_b,\,\omega_c)$, each taken at their $\pm 95\%$~C.L. \textit{Planck}~2018 limits, relative to the fiducial  (\textit{Planck}~2018 best-fit) cosmology used in \texttt{DarkHistory}. As a representative example, we take $m_{\mathrm{DM}} = 10\,\mathrm{keV}$ and consider three choices of scalar DM-electron coupling $g_{\phi \bar{e}e}$: the first yields a DM lifetime at the 95\%~C.L. exclusion limit reported in Ref.~\cite{2024arXiv240813305X}, while the other two are an order of magnitude larger and smaller than this baseline value. Equivalent plots for the other DM masses relevant to our analysis are provided for completeness in Appendix~\ref{app:fs_cosmo}. Overall, we observe $\sim \mathcal{O}(1)\%$ residual changes in $f_{\rm H_{ion}}(z)$, which remain roughly independent of $g_{\phi \bar{e}e}$ near recombination but grow slightly larger at lower redshifts for stronger couplings. These results support the expectation that cosmology-driven variations in $f_c(z)$ are modest~\cite{Liu:2019bbm}, yet they motivate a dedicated check to ensure that such percent-level corrections to the $f_c(z)$ do not bias the final DM-electron coupling limits derived from a full MCMC treatment. This verification has not been explicitly performed in previous studies, and we address it in the next section.

\section{Analysis}
\label{sec:analysis}

For the three models of interest in Sec.~\ref{sec:models}, we perform MCMC analyses to place constraints on DM-electron coupling in the DM mass range between 100~eV and 1~MeV using \textit{Planck} 2018 temperature, polarization, and lensing data.
Since investigations into the effects of decaying DM have been performed previously in the literature~\cite{Liu:2023nct, Capozzi:2023xie, 2024arXiv240813305X}, we focus on two complementary aspects: \textit{(i)} incorporating a cosmology-consistent treatment of energy injection and \textit{(ii)} leveraging the improved precision of the latest \textit{Planck} PR4 data release.

Regarding aspect~\textit{(i)}, as we discuss in Sec.~\ref{sec:energy_injection}, previous studies using \texttt{DarkHistory} assume a fiducial cosmology when computing the energy deposition functions $f_c(z)$, independently from any variation of the other sampled cosmological parameters in the MCMC. We test for potential bias that could arise from fixing the cosmology.

Regarding aspect~\textit{(ii)}, the recent \texttt{NPIPE} maps offer advancements compared to previous \textit{Planck} data releases, including increased sky coverage at high frequencies and improved processing of time-ordered data, which lead to a reduction in small-scale noise.
These improvements provide 10\% to 20\% smaller uncertainties on constraints of~$\Lambda$CDM parameters and minimal extensions~\cite{Tristram:2023haj}, further motivating their application to decaying DM models.

In order to address these aspects, we perform the following three CMB analyses:
\begin{description}
\item[PR3-$f^\mathrm{FC}_c$]
  We use the~\textit{Planck} 2018 PR3 likelihood code, employing the \texttt{plik-lite} likelihood for high multipoles ($\ell \geq 30$), the~\texttt{commander} and~\texttt{SimAll} likelihoods at low multipoles ($2 \leq \ell \leq 29$) for the temperature and polarization power spectra, respectively, and the~\texttt{SMICA} lensing reconstruction likelihood~\cite{Planck:2019nip}.
  The~$f_c(z)$ functions are precomputed by~\texttt{DarkHistory} for different values of masses and couplings under the assumption of a fixed fiducial cosmology (FC), and we interpolate for intermediate values of the couplings.
\item[PR3-$f^\mathrm{CC}_c$]
  We use the~\textit{Planck} 2018 PR3 likelihood code as described for~\textbf{PR3-$f^\mathrm{FC}_c$}.
  However, this analysis incorporates a cosmology-consistent (CC) treatment of energy injection by recomputing the~$f_c(z)$ functions at each step of the MCMC, i.e., adjusting~\texttt{DarkHistory} cosmological parameters to the current sampled value of the chain.
\item[PR4-$f^\mathrm{FC}_c$]
  We use the latest~\texttt{NPIPE} \textit{Planck} data release and use the corresponding~\textit{Planck} PR4 likelihood code~\cite{Planck:2020olo}.
  We replace the \textit{Planck} PR3 \texttt{plik-lite} high-$\ell$ likelihood with the updated~\texttt{HiLLiPoP} likelihood~\cite{Tristram:2023haj}.
  For the polarization spectra at~$\ell < 30$ we use the PR4~\texttt{LoLLiPoP} likelihood~\cite{Tristram:2023haj} instead of the PR3 \texttt{SimAll} likelihood, while retaining the low-$\ell$ \texttt{Commander} likelihood for temperature power spectra and the~\texttt{SMICA} lensing reconstruction likelihood.\footnote{An updated PR4 lensing likelihood based on the~\texttt{NPIPE} data release is also available~\cite{Carron:2022eyg}, but we do not include it in our analysis. Decaying DM has negligible impact on the lensing convergence signal when the decay time is significantly longer than the age of the universe~\cite{Li:2018zdm}. Incorporating this likelihood would only marginally tighten constraints on the DM density, with no appreciable effects on our exclusions limits.}
  As described for \textbf{PR3-$f^\mathrm{FC}_c$}, the~$f_c(z)$ functions are precomputed by~\texttt{DarkHistory} under a fixed fiducial cosmology.
\end{description}

\begin{figure*}[ht]
    \centering
    \includegraphics[width=\linewidth]{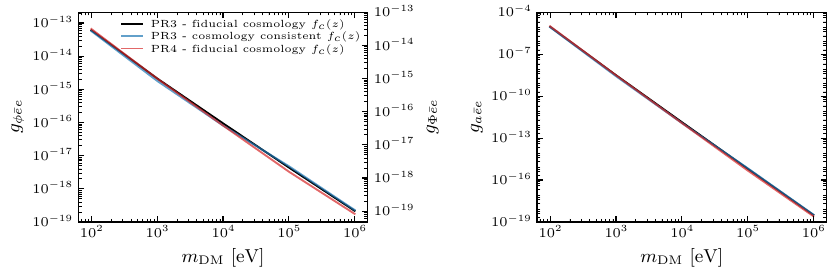}
    \caption{\textit{Planck} 95\% C.L. upper limits on the dimensionless DM-electron coupling in the scalar ($g_{\phi \bar{e}e}$) (\textit{left panel, left axis}), pseudo-scalar ($g_{\Phi \bar{e}e}$) (\textit{left panel, right axis}), and ALP ($g_{a \bar{e}e}$) (\textit{right panel}) cases as a function of the DM mass, $m_{\rm{DM}}$. 
    For each panel, the black curve shows the results from the \textit{Planck} 2018 PR3 likelihood with the energy deposition functions $f_c(z)$ from Eq.~\eqref{eq:f_c} pre-computed under a fiducial $\Lambda$CDM cosmology (\textbf{PR3-$f_c^{\rm{FC}}$}); the \textcolor{tabblue}{blue} curve corresponds to the PR3 analysis for which the $f_c(z)$ functions are evaluated at each MCMC step to reflect the sampled cosmological parameters (\textbf{PR3-$f_c^{\rm{CC}}$}); and the \textcolor{tabred}{red} curve corresponds to the constraints obtained with the updated \textit{Planck} PR4 likelihood using the precomputed $f_c(z)$ functions from a fiducial cosmology (\textbf{PR4-$f_c^{\rm{FC}}$}).}
\label{fig:CMB_limits}
\end{figure*}

For each dataset, we sample the following cosmological parameters with broad flat priors: the angular size of the sound horizon at decoupling~$\theta_s$, the physical baryon density~$\omega_b$, the physical matter density~$\omega_m$, the scalar amplitude~$\ln(10^{10}A_s)$, the scalar spectral index~$n_s$, and the redshift at reionization $z_{\rm{reio}}$.
Additionally, we sample~$\log_{10}\left( g_{\phi \bar{e}e}\right)$ with a flat prior for fixed DM masses $m_\mathrm{DM}=\{100\,\rm{eV},\,1\,\rm{keV},\,10\,\rm{keV},\,100\,\rm{keV},\,1\,\rm{MeV}\}$.
Since the structures of the pseudo-scalar and ALP decay widths are formally equivalent to that of scalar DM, it is possible to recast scalar DM upper limits into pseudo-scalar and ALP limits via the rescalings~$g_{\phi \bar{e}e} \rightarrow \sqrt{288/64}\,g_{\Phi\bar{e}e}$ and~$g_{\phi \bar{e}e}\rightarrow \sqrt{288/64}\,m_{\rm{DM}}^2/(12\,m_e^2) \,g_{a\bar{e}e}$, respectively.

We interface \texttt{CLASS v3.2.5} with the publicly available sampler \texttt{MontePython}~\cite{Audren:2012wb, Brinckmann:2018cvx} to perform the MCMC analyses.
Chain convergence is reached when the Gelman-Rubin criterion~\cite{Gelman:1992zz} $R < 0.01$ is satisfied.
Finally, we utilize \texttt{GetDist}~\cite{2019arXiv191013970L} to process our MCMC chains.

Due to the high computational cost of the~\textbf{PR3-$f^\mathrm{CC}_c$} analysis, we limit~\texttt{DarkHistory} to~$n = 10$ hydrogen levels. 
This adjustment significantly reduces runtime while remaining a valid approximation for DM masses well above the hydrogen ionization energy~\cite{Liu:2023fgu, Liu:2023nct}.
We impose the same limit of $n=10$ levels for the other two analyses to maintain consistency with \textbf{PR3-$f^\mathrm{CC}_c$}.

\section{Results}
\label{sec:results}

We present the results of our analyses in Fig.~\ref{fig:CMB_limits} using the \textit{Planck} 2018 PR3 likelihoods with cosmology-consistent energy deposition functions (\textbf{PR3-$f_c^{\rm{CC}}$}) in \textcolor{tabblue}{blue}, the PR3 likelihoods with precomputed energy deposition functions (\textbf{PR3-$f_c^{\rm{FC}}$}) in black, and the PR4 likelihoods with precomputed energy deposition functions (\textbf{PR4-$f_c^{\rm{FC}}$}) in \textcolor{tabred}{red}.
For each analysis choice, we show the 95\% confidence level (C.L.) upper limits on the dimensionless DM coupling constants~$g_{\phi \bar{e}e}$, $g_{\Phi \bar{e}e}$, and~$g_{a \bar{e}e}$ as a function of the DM mass for the scalar (left vertical axis of the left panel), pseudo-scalar (right vertical axis of the left panel), and ALP (right panel) cases, respectively.

As expected, our results from the standard, fixed-cosmology \textbf{PR3-$f_c^{\rm{FC}}$} analysis are consistent with the bounds on DM decay reported in previous studies~\cite{Capozzi:2023xie, Liu:2023nct, 2024arXiv240813305X}.
Importantly, our results for the cosmology-consistent \textbf{PR3-$f_c^{\rm{CC}}$} analysis aligns with those from fixed-cosmology approach at the 10\% level.
This agreement validates the standard methodology of assuming a fixed fiducial $\Lambda$CDM cosmology when pre-computing the~$f_c(z)$ functions from \texttt{DarkHistory} for decaying DM.
It further highlights that percent-level differences in the computation of the energy deposition functions do not have any appreciable impact on our DM decay bounds. As discussed in Appendix~\ref{app:MCMC_analysis}, we observe that the cosmology-consistent approach reduces mild parameter degeneracies present in the fixed-cosmology case, particularly at low DM masses, though this reduction has negligible impact on the derived constraints.

We find that the upper limits derived from the \textit{Planck} PR4 analysis are consistent with those obtained using the PR3 likelihoods, with only minor differences observed for $m_{\rm{DM}}\geq 10\,$keV. 
These~$20\%$ improvements in the DM-electron coupling constant exclusion limits can be attributed to lower noise levels in the~\texttt{LoLLiPoP} likelihood used for low-$\ell$ polarization spectra in PR4.
However, for $m_{\rm{DM}} \lesssim 10\,$keV, the constraints are effectively the same, likely due to increased degeneracy between $g_{\phi\bar{e}e}$, $\ln (10^{10} A_s)$, and $z_{\rm{reio}}$ in this mass range, which further limits the impact of the improved polarization data (see also Appendix~\ref{app:MCMC_analysis}).
In general, the lack of significant improvements in the upper limits reflects the fact that most of the constraining power for decaying DM originates from low-$\ell$ polarization data, where the differences between PR3 and PR4 are modest, limiting the potential gains from the enhanced high-$\ell$ precision in the~\texttt{NPIPE} data release.


\subsection{Flavorful Variations}

Our framework also applies to the cases of 1-loop muon- and tau-mediated decays; therefore, models in which DM couples exclusively to muons or taus are also constrained by our analysis. Such DM models are well-motivated ~\cite{Ziegler:2019gjr, PANCI2023137919, Ziegler:2023aoe} and open up a host of collider~\cite{Bauer:2021mvw, 2025arXiv250106294L} and astrophysical probes~\cite{Bollig:2020xdr, Croon:2020lrf, Green:2021hjh, Caputo:2021rux}. 
The fact that such models are constrained by the CMB even though they do not couple to electrons (i.e., no additional tree-level scatterings processes are present at all during the epoch of recombination) further highlights the importance of the 1-loop decay.
Decay widths in these scenario are obtained with a simple rescaling~$m_e \rightarrow m_{\mu}$ or~$ m_{\tau}$ in Eqs.~\eqref{eq:Gamma_scalar}, \eqref{eq:Gamma_pseudoscalar} and~\eqref{eq:Gamma_alp}. 
In other words, the bounds presented in Fig.~\ref{fig:CMB_limits} for the DM-electron coupling equally apply to the DM-muon or DM-tau coupling scenarios, after rescaling them by a factor~$m_\mu/m_e \sim 10^3$ and~$m_\tau/m_e\sim 10^{5}$, respectively.

More generally, leptophilic DM models with couplings to all three flavors decay at a rate give by
\begin{equation}
    \Gamma_{\phi \rightarrow \gamma \gamma} = \frac{\alpha_{\rm{EM}}^2}{288 \pi^3} \sum_{i,j = e,\mu,\tau}  g_{\phi \bar l_i l_j}^2  \frac{m_\phi^3}{m_{l_i} m_{l_j}}\,,
\end{equation}
with an analogous expressions for pseudo-scalar DM. 
For universal flavor couplings, the contribution from muon and tau leptons in the loop have a negligible effect on our bound in Fig.~\ref{fig:CMB_limits}. 
Finally, in agreement with the literature~\cite{PANCI2023137919}, we find that CMB constraints do not shed light on the parameter space of interest to lepton flavor violating decays.

\section{Comparison with other constraints}
\label{sec:comparison}

\begin{figure}[t]
    \centering
    \includegraphics[width=1\linewidth]{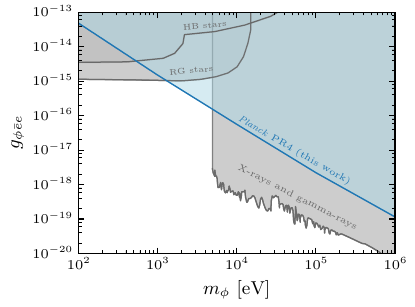}
    \caption{Comparison of our \textit{Planck} $95\%$ C.L.\ CMB upper limit on the scalar DM-electron coupling $g_{\phi \bar{e}e}$ (in \textcolor{tabblue}{blue}) with limits from previous studies.
    We show (in \textcolor{tabgray}{gray}) stellar cooling constraints from horizontal branch (HB) and red giant (RG) stars~\cite{Hardy:2016kme}, as well as X-ray and gamma-ray constraints from NuStar, XMM-Newton, and INTEGRAL data~\cite{Ng:2019gch,Laha:2020ivk,Foster:2021ngm,Ferreira:2022egk}.}
    \label{fig:Scalar}
\end{figure}

\begin{figure}[t]
    \centering
    \includegraphics[width=1\linewidth]{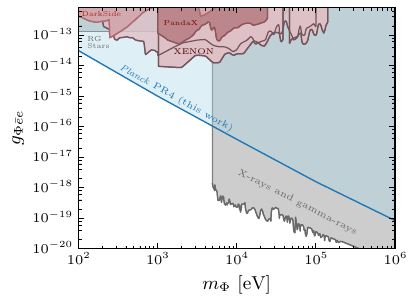}
    \caption{Comparison of our \textit{Planck} $95\%$ C.L.\ CMB upper limit on the pseudo-scalar DM-electron coupling $g_{\Phi \bar{e}e}$ (in \textcolor{tabblue}{blue}) with limits from previous studies. We show (in \textcolor{tabgray}{gray}) stellar cooling constraints from red giant (RG) stars~\cite{Hardy:2016kme, Capozzi:2020cbu}, as well as X-ray and gamma-ray constraints from NuStar, XMM-Newton, and INTEGRAL data~\cite{Ng:2019gch,Laha:2020ivk,Foster:2021ngm,Ferreira:2022egk}. We also show (in different shades of \textcolor{tabred}{red}) direct detection bounds from from DarkSide~\cite{DarkSide:2022knj}, PandaX~\cite{PandaX:2024cic,PandaX:2024kjp}, and XENON~\cite{XENON:2019gfn,XENON:2020rca,XENON:2021myl,XENONCollaboration:2022kmb}.}
    \label{fig:Pseudoscalar}
\end{figure}

\begin{figure}[t]
    \centering
    \includegraphics[width=1\linewidth]{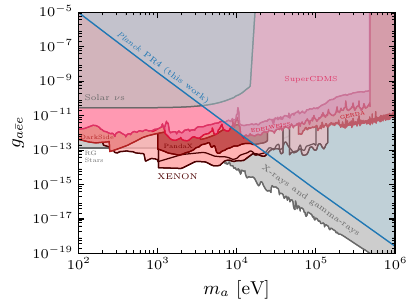}
    \caption{Comparison of our \textit{Planck} $95\%$ C.L.\ CMB upper limits on the ALP-electron coupling $g_{a \bar{e}e}$ (in \textcolor{tabblue}{blue}) with limits from previous studies.
    We show (in \textcolor{tabgray}{gray}) stellar cooling constraints from red giant (RG) stars~\cite{Hardy:2016kme, Capozzi:2020cbu}, the solar luminosity/neutrino bound~\cite{Croon:2020lrf}, and X-ray and gamma-ray constraints from NuStar, XMM-Newton, and INTEGRAL~\cite{Ng:2019gch,Laha:2020ivk,Foster:2021ngm,Ferreira:2022egk}.
    We also show (in different shades of \textcolor{tabred}{red}) direct detection bounds from DarkSide~\cite{DarkSide:2022knj}, PandaX~\cite{PandaX:2024cic,PandaX:2024kjp}, SuperCDMS~\cite{Aralis:2019nfa}, EDELWEISS~\cite{EDELWEISS:2018tde}, GERDA~\cite{GERDA:2020emj}, and XENON~\cite{XENON:2019gfn,XENON:2020rca,XENON:2021myl,XENONCollaboration:2022kmb}.}
    \label{fig:Axion}
\end{figure}

In Figs.~\ref{fig:Scalar}, \ref{fig:Pseudoscalar}, and~\ref{fig:Axion}, we show our strongest upper limits on scalar, pure pseudo-scalar, and ALP DM-electron couplings from our \textit{Planck} PR4 analysis and compare them with existing bounds in the literature.

For the scalar case shown in Fig.~\ref{fig:Scalar}, stellar cooling constraints from horizontal branch (HB) and red giants (RG) stars~\cite{Hardy:2016kme,Bottaro:2023gep} provided the most stringent bounds on~$g_{\phi\bar{e} e}$ for~$m_\phi\lesssim 50\,$keV prior to our work. For masses larger than~$50$ keV, we re-cast constraints on the non-detection of X-ray and gamma-ray signals from the NuStar, XMM-Newton, and INTEGRAL telescopes~\cite{Ng:2019gch,Laha:2020ivk,Foster:2021ngm,Ferreira:2022egk} produced by $\phi \to \gamma \gamma$ decays.
Exclusion limits from CMB anisotropies derived in this work, shown in \textcolor{tabblue}{blue}, surpass previous bounds for $1\,\text{keV}\lesssim m_{\phi} \lesssim 5\,\text{keV}$ by up to an order of magnitude in this range.
Although our CMB bounds are weaker than the astrophysical bounds in most of the mass range shown, they are robust and are not subject to the same astrophysical uncertainties.

In the case of pseudo-scalars and ALPs shown in Figs.~\ref{fig:Pseudoscalar} and~\ref{fig:Axion}, respectively, DM-electron interactions are equivalent at tree-level. Consequently, terrestrial limits and stellar constraints are identical for the two classes of models, although we display only a subset of these bounds in each figure to highlight the parameter space most relevant for our derived CMB constraints.  Specifically, for the pure pseudo-scalar case in Fig.~\ref{fig:Pseudoscalar}, we show the direct detection bounds from DarkSide~\cite{DarkSide:2022knj}, PandaX~\cite{PandaX:2024cic,PandaX:2024kjp}
and XENON~\cite{XENON:2019gfn, XENON:2020rca, XENON:2021myl, XENONCollaboration:2022kmb}, along with the stellar cooling constraints from RG stars~\cite{Hardy:2016kme, Capozzi:2020cbu}. For ALPs~in Fig.~\ref{fig:Axion}, we also include the SuperCDMS~\cite{Aralis:2019nfa}, EDELWEISS~\cite{EDELWEISS:2018tde}, and GERDA~\cite{GERDA:2020emj} exclusion limits,
as well as
the solar luminosity/neutrino bound~\cite{Croon:2020lrf}.

In contrast to their identical behavior at tree level, at 1-loop the ALP decay width scales as $\Gamma_{a\to\gamma\gamma} \sim (m_a/m_e)^6$, whereas the pseudo-scalar decay width scales as $\Gamma_{\Phi\to\gamma\gamma} \sim (m_\Phi/m_e)^2$.
Thus, the X-ray, gamma-ray, and CMB upper bounds are significantly weaker for ALPs.
Despite this difference, the X-ray and gamma-ray constraints consistently provide the leading bounds for both pseudo-scalar and ALP DM for $m_\text{DM} \gtrsim 5\,$keV. 
On the other hand, for~$m_\text{DM} \lesssim 5\,$keV, CMB bounds on pseudo-scalar DM-electron coupling surpass all existing upper bounds, exceeding the XENON limits up to two orders of magnitude at $m_\Phi \sim 1\,$keV and the RG bound by roughly an order of magnitude for $m_\Phi \sim 100\,$eV.
Conversely, the steep mass-dependent scaling causes the CMB constraints on ALP-electron coupling to remain subdominant to both terrestrial and astrophysical bounds. 

\section{Conclusions}
\label{sec:conclusions}

In this work, we study constraints from CMB anisotropies on sub-MeV spin-zero DM, focusing on scalar, pseudo-scalar, and ALP models that couple leptophilically to the Standard Model. 
These scenarios predict DM decay into photons at the 1-loop level, which is the dominant process for energy injection in the early universe.

We expand upon previous CMB studies of DM decay~\cite{Liu:2023nct, Capozzi:2023xie, 2024arXiv240813305X} by conducting two complementary MCMC analyses. 
First, we explore a cosmology-consistent treatment of energy injection, dynamically computing the $f_c(z)$ functions at each MCMC step to reflect variations in cosmological parameters, rather than relying on precomputed profiles based on a fixed fiducial cosmology. 
Second, we utilize the improved precision of the \textit{Planck} PR4 likelihoods~\cite{Tristram:2023haj}, based on the \texttt{NPIPE}~\cite{Planck:2020olo} data release, to examine their impact on the derived constraints.

We find that the cosmology-consistent analysis reproduces the results obtained using the standard fixed-cosmology approach, validating the latter methodology for deriving constraints on DM decay.
This outcome is important for future studies, since running a cosmology-consistent analysis with \texttt{DarkHistory} takes significantly more computational time and resources than running a fixed-cosmology analysis.
Our results using the \textit{Planck} PR3 and PR4 likelihoods are largely consistent, with only minor improvements observed for $m_{\rm{DM}} \gtrsim 10\,$keV. 
This consistency reflects the fact that constraints on decaying DM are predominantly driven by low-$\ell$ polarization data, where the improvements introduced by the PR4 likelihoods are modest. 

We compare our derived constraints with existing bounds from astrophysical sources and terrestrial experiments. 
For scalar DM, our CMB limits surpass previous bounds for $1\,{\rm keV} \lesssim m_{\phi} \lesssim 5\,$keV by up to an order of magnitude. For pseudo-scalar DM, the CMB constraints provide the leading limit for $m_{\Phi} \lesssim 5\,$keV, exceeding terrestrial constraints, specifically those from the XENON experiment~\cite{XENON:2019gfn, XENON:2020rca, XENON:2021myl, XENONCollaboration:2022kmb}, by up to two orders of magnitude. 
In the ALP case, terrestrial and astrophysical searches still provide the dominant limits across our mass range of interest.

Since our results arise from electron loop-level decays into photons, similar bound on DM-lepton coupling arises in leptophilic spin-zero DM models that interact exclusively to muons or taus (rescaled by $m_e \rightarrow m_\mu$ or $m_\tau$). 
We note that for leptophilic vector or pseudo-vector DM models, the loop-induced decay to two photons is forbidden by charge conjugation invariance. We explore the CMB bounds on these models in upcoming work.

As a whole, our findings underscore the powerful role of CMB anisotropies in probing exotic energy injection mechanisms, providing a robust and powerful complement to astrophysical and terrestrial constraints. 


\begin{acknowledgments}
We are grateful to Can Kilic, Hongwan Liu and Wenzer Qin for useful discussions.
We acknowledge the use of \texttt{CLASS}~\cite{Blas:2011rf}, \texttt{GetDist}~\cite{2019arXiv191013970L}, \texttt{IPython}~\cite{Perez:2007ipy}, \texttt{MontePython}~\cite{Audren:2012wb, Brinckmann:2018cvx}, and the Python packages \texttt{Matplotlib}~\cite{Hunter:2007mat}, \texttt{NumPy}~\cite{Harris:2020xlr} and~\texttt{SciPy}~\cite{Virtanen:2019joe}. 
We also acknowledge the use of the \texttt{AxionLimits} Github repository~\cite{AxionLimits},which was an invaluable resource in verifying existing axion constraints and extracting most of the bounds from the current literature shown in Figs.~\ref{fig:Pseudoscalar} and~\ref{fig:Axion}.

We acknowledge the Texas Advanced Computing Center (TACC) at The University of
Texas at Austin for providing high-performance computing resources that have contributed to the research results reported within this paper.

NB is supported by PRD/ARPE 2022 ``Cosmology with Gravitational waves and Large Scale Structure - CosmoGraLSS''.
KB acknowledges support from the National Science Foundation under Grant No.~PHY-2413016.
GM acknowledges support by the U.S.\ Department of Energy, Office of Science, Office of High Energy Physics program under Award Number DE-SC-0022021, as well as the Swedish Research Council (Contract No.~638-2013-8993). GM is also supported by the Continuing Fellowship of the Graduate School of the College of Natural Sciences at the University of Texas at Austin. 
\end{acknowledgments}


\bibliography{references.bib}

\begin{thebibliography}{86}%
\makeatletter
\providecommand \@ifxundefined [1]{%
 \@ifx{#1\undefined}
}%
\providecommand \@ifnum [1]{%
 \ifnum #1\expandafter \@firstoftwo
 \else \expandafter \@secondoftwo
 \fi
}%
\providecommand \@ifx [1]{%
 \ifx #1\expandafter \@firstoftwo
 \else \expandafter \@secondoftwo
 \fi
}%
\providecommand \natexlab [1]{#1}%
\providecommand \enquote  [1]{``#1''}%
\providecommand \bibnamefont  [1]{#1}%
\providecommand \bibfnamefont [1]{#1}%
\providecommand \citenamefont [1]{#1}%
\providecommand \href@noop [0]{\@secondoftwo}%
\providecommand \href [0]{\begingroup \@sanitize@url \@href}%
\providecommand \@href[1]{\@@startlink{#1}\@@href}%
\providecommand \@@href[1]{\endgroup#1\@@endlink}%
\providecommand \@sanitize@url [0]{\catcode `\\12\catcode `\$12\catcode `\&12\catcode `\#12\catcode `\^12\catcode `\_12\catcode `\%12\relax}%
\providecommand \@@startlink[1]{}%
\providecommand \@@endlink[0]{}%
\providecommand \url  [0]{\begingroup\@sanitize@url \@url }%
\providecommand \@url [1]{\endgroup\@href {#1}{\urlprefix }}%
\providecommand \urlprefix  [0]{URL }%
\providecommand \Eprint [0]{\href }%
\providecommand \doibase [0]{https://doi.org/}%
\providecommand \selectlanguage [0]{\@gobble}%
\providecommand \bibinfo  [0]{\@secondoftwo}%
\providecommand \bibfield  [0]{\@secondoftwo}%
\providecommand \translation [1]{[#1]}%
\providecommand \BibitemOpen [0]{}%
\providecommand \bibitemStop [0]{}%
\providecommand \bibitemNoStop [0]{.\EOS\space}%
\providecommand \EOS [0]{\spacefactor3000\relax}%
\providecommand \BibitemShut  [1]{\csname bibitem#1\endcsname}%
\let\auto@bib@innerbib\@empty
\bibitem [{\citenamefont {Weinberg}(1978)}]{PhysRevLett.40.223}%
  \BibitemOpen
  \bibfield  {author} {\bibinfo {author} {\bibfnamefont {S.}~\bibnamefont {Weinberg}},\ }\bibfield  {title} {\bibinfo {title} {A new light boson?},\ }\href {https://doi.org/10.1103/PhysRevLett.40.223} {\bibfield  {journal} {\bibinfo  {journal} {Phys. Rev. Lett.}\ }\textbf {\bibinfo {volume} {40}},\ \bibinfo {pages} {223} (\bibinfo {year} {1978})}\BibitemShut {NoStop}%
\bibitem [{\citenamefont {Adams}\ \emph {et~al.}(2022)\citenamefont {Adams} \emph {et~al.}}]{Adams:2022pbo}%
  \BibitemOpen
  \bibfield  {author} {\bibinfo {author} {\bibfnamefont {C.~B.}\ \bibnamefont {Adams}} \emph {et~al.},\ }\bibfield  {title} {\bibinfo {title} {{Axion Dark Matter}},\ }in\ \href@noop {} {\emph {\bibinfo {booktitle} {{Snowmass 2021}}}}\ (\bibinfo {year} {2022})\ \Eprint {https://arxiv.org/abs/2203.14923} {arXiv:2203.14923 [hep-ex]} \BibitemShut {NoStop}%
\bibitem [{\citenamefont {Choi}\ \emph {et~al.}(2021)\citenamefont {Choi}, \citenamefont {Im},\ and\ \citenamefont {Sub~Shin}}]{Choi:2020rgn}%
  \BibitemOpen
  \bibfield  {author} {\bibinfo {author} {\bibfnamefont {K.}~\bibnamefont {Choi}}, \bibinfo {author} {\bibfnamefont {S.~H.}\ \bibnamefont {Im}},\ and\ \bibinfo {author} {\bibfnamefont {C.}~\bibnamefont {Sub~Shin}},\ }\bibfield  {title} {\bibinfo {title} {{Recent Progress in the Physics of Axions and Axion-Like Particles}},\ }\href {https://doi.org/10.1146/annurev-nucl-120720-031147} {\bibfield  {journal} {\bibinfo  {journal} {Ann. Rev. Nucl. Part. Sci.}\ }\textbf {\bibinfo {volume} {71}},\ \bibinfo {pages} {225} (\bibinfo {year} {2021})},\ \Eprint {https://arxiv.org/abs/2012.05029} {arXiv:2012.05029 [hep-ph]} \BibitemShut {NoStop}%
\bibitem [{\citenamefont {Raffelt}(2007)}]{Raffelt:2006rj}%
  \BibitemOpen
  \bibfield  {author} {\bibinfo {author} {\bibfnamefont {G.~G.}\ \bibnamefont {Raffelt}},\ }\bibfield  {title} {\bibinfo {title} {{Axions: Motivation, limits and searches}},\ }\href {https://doi.org/10.1088/1751-8113/40/25/S05} {\bibfield  {journal} {\bibinfo  {journal} {J. Phys. A}\ }\textbf {\bibinfo {volume} {40}},\ \bibinfo {pages} {6607} (\bibinfo {year} {2007})},\ \Eprint {https://arxiv.org/abs/hep-ph/0611118} {arXiv:hep-ph/0611118} \BibitemShut {NoStop}%
\bibitem [{\citenamefont {Batell}\ \emph {et~al.}(2024)\citenamefont {Batell}, \citenamefont {Ghalsasi},\ and\ \citenamefont {Rai}}]{Batell:2022qvr}%
  \BibitemOpen
  \bibfield  {author} {\bibinfo {author} {\bibfnamefont {B.}~\bibnamefont {Batell}}, \bibinfo {author} {\bibfnamefont {A.}~\bibnamefont {Ghalsasi}},\ and\ \bibinfo {author} {\bibfnamefont {M.}~\bibnamefont {Rai}},\ }\bibfield  {title} {\bibinfo {title} {{Dynamics of dark matter misalignment through the Higgs portal}},\ }\href {https://doi.org/10.1007/JHEP01(2024)038} {\bibfield  {journal} {\bibinfo  {journal} {JHEP}\ }\textbf {\bibinfo {volume} {01}},\ \bibinfo {pages} {038}},\ \Eprint {https://arxiv.org/abs/2211.09132} {arXiv:2211.09132 [hep-ph]} \BibitemShut {NoStop}%
\bibitem [{\citenamefont {Preskill}\ \emph {et~al.}(1983)\citenamefont {Preskill}, \citenamefont {Wise},\ and\ \citenamefont {Wilczek}}]{PRESKILL1983127}%
  \BibitemOpen
  \bibfield  {author} {\bibinfo {author} {\bibfnamefont {J.}~\bibnamefont {Preskill}}, \bibinfo {author} {\bibfnamefont {M.~B.}\ \bibnamefont {Wise}},\ and\ \bibinfo {author} {\bibfnamefont {F.}~\bibnamefont {Wilczek}},\ }\bibfield  {title} {\bibinfo {title} {Cosmology of the invisible axion},\ }\href {https://doi.org/https://doi.org/10.1016/0370-2693(83)90637-8} {\bibfield  {journal} {\bibinfo  {journal} {Physics Letters B}\ }\textbf {\bibinfo {volume} {120}},\ \bibinfo {pages} {127} (\bibinfo {year} {1983})}\BibitemShut {NoStop}%
\bibitem [{\citenamefont {Dine}\ and\ \citenamefont {Fischler}(1983)}]{DINE1983137}%
  \BibitemOpen
  \bibfield  {author} {\bibinfo {author} {\bibfnamefont {M.}~\bibnamefont {Dine}}\ and\ \bibinfo {author} {\bibfnamefont {W.}~\bibnamefont {Fischler}},\ }\bibfield  {title} {\bibinfo {title} {The not-so-harmless axion},\ }\href {https://doi.org/https://doi.org/10.1016/0370-2693(83)90639-1} {\bibfield  {journal} {\bibinfo  {journal} {Physics Letters B}\ }\textbf {\bibinfo {volume} {120}},\ \bibinfo {pages} {137} (\bibinfo {year} {1983})}\BibitemShut {NoStop}%
\bibitem [{\citenamefont {Davis}(1986)}]{DAVIS1986225}%
  \BibitemOpen
  \bibfield  {author} {\bibinfo {author} {\bibfnamefont {R.}~\bibnamefont {Davis}},\ }\bibfield  {title} {\bibinfo {title} {Cosmic axions from cosmic strings},\ }\href {https://doi.org/https://doi.org/10.1016/0370-2693(86)90300-X} {\bibfield  {journal} {\bibinfo  {journal} {Physics Letters B}\ }\textbf {\bibinfo {volume} {180}},\ \bibinfo {pages} {225} (\bibinfo {year} {1986})}\BibitemShut {NoStop}%
\bibitem [{\citenamefont {Gorghetto}\ \emph {et~al.}(2018)\citenamefont {Gorghetto}, \citenamefont {Hardy},\ and\ \citenamefont {Villadoro}}]{Gorghetto:2018myk}%
  \BibitemOpen
  \bibfield  {author} {\bibinfo {author} {\bibfnamefont {M.}~\bibnamefont {Gorghetto}}, \bibinfo {author} {\bibfnamefont {E.}~\bibnamefont {Hardy}},\ and\ \bibinfo {author} {\bibfnamefont {G.}~\bibnamefont {Villadoro}},\ }\bibfield  {title} {\bibinfo {title} {{Axions from Strings: the Attractive Solution}},\ }\href {https://doi.org/10.1007/JHEP07(2018)151} {\bibfield  {journal} {\bibinfo  {journal} {JHEP}\ }\textbf {\bibinfo {volume} {07}},\ \bibinfo {pages} {151}},\ \Eprint {https://arxiv.org/abs/1806.04677} {arXiv:1806.04677 [hep-ph]} \BibitemShut {NoStop}%
\bibitem [{\citenamefont {Gorghetto}\ \emph {et~al.}(2021)\citenamefont {Gorghetto}, \citenamefont {Hardy},\ and\ \citenamefont {Villadoro}}]{Gorghetto:2020qws}%
  \BibitemOpen
  \bibfield  {author} {\bibinfo {author} {\bibfnamefont {M.}~\bibnamefont {Gorghetto}}, \bibinfo {author} {\bibfnamefont {E.}~\bibnamefont {Hardy}},\ and\ \bibinfo {author} {\bibfnamefont {G.}~\bibnamefont {Villadoro}},\ }\bibfield  {title} {\bibinfo {title} {{More axions from strings}},\ }\href {https://doi.org/10.21468/SciPostPhys.10.2.050} {\bibfield  {journal} {\bibinfo  {journal} {SciPost Phys.}\ }\textbf {\bibinfo {volume} {10}},\ \bibinfo {pages} {050} (\bibinfo {year} {2021})},\ \Eprint {https://arxiv.org/abs/2007.04990} {arXiv:2007.04990 [hep-ph]} \BibitemShut {NoStop}%
\bibitem [{\citenamefont {Lyth}(1992)}]{Lyth:1991ub}%
  \BibitemOpen
  \bibfield  {author} {\bibinfo {author} {\bibfnamefont {D.~H.}\ \bibnamefont {Lyth}},\ }\bibfield  {title} {\bibinfo {title} {{Axions and inflation: Sitting in the vacuum}},\ }\href {https://doi.org/10.1103/PhysRevD.45.3394} {\bibfield  {journal} {\bibinfo  {journal} {Phys. Rev. D}\ }\textbf {\bibinfo {volume} {45}},\ \bibinfo {pages} {3394} (\bibinfo {year} {1992})}\BibitemShut {NoStop}%
\bibitem [{\citenamefont {Co}\ \emph {et~al.}(2020{\natexlab{a}})\citenamefont {Co}, \citenamefont {Hall},\ and\ \citenamefont {Harigaya}}]{Co:2019jts}%
  \BibitemOpen
  \bibfield  {author} {\bibinfo {author} {\bibfnamefont {R.~T.}\ \bibnamefont {Co}}, \bibinfo {author} {\bibfnamefont {L.~J.}\ \bibnamefont {Hall}},\ and\ \bibinfo {author} {\bibfnamefont {K.}~\bibnamefont {Harigaya}},\ }\bibfield  {title} {\bibinfo {title} {{Axion Kinetic Misalignment Mechanism}},\ }\href {https://doi.org/10.1103/PhysRevLett.124.251802} {\bibfield  {journal} {\bibinfo  {journal} {Phys. Rev. Lett.}\ }\textbf {\bibinfo {volume} {124}},\ \bibinfo {pages} {251802} (\bibinfo {year} {2020}{\natexlab{a}})},\ \Eprint {https://arxiv.org/abs/1910.14152} {arXiv:1910.14152 [hep-ph]} \BibitemShut {NoStop}%
\bibitem [{\citenamefont {Co}\ \emph {et~al.}(2019)\citenamefont {Co}, \citenamefont {Gonzalez},\ and\ \citenamefont {Harigaya}}]{Co:2018mho}%
  \BibitemOpen
  \bibfield  {author} {\bibinfo {author} {\bibfnamefont {R.~T.}\ \bibnamefont {Co}}, \bibinfo {author} {\bibfnamefont {E.}~\bibnamefont {Gonzalez}},\ and\ \bibinfo {author} {\bibfnamefont {K.}~\bibnamefont {Harigaya}},\ }\bibfield  {title} {\bibinfo {title} {{Axion Misalignment Driven to the Hilltop}},\ }\href {https://doi.org/10.1007/JHEP05(2019)163} {\bibfield  {journal} {\bibinfo  {journal} {JHEP}\ }\textbf {\bibinfo {volume} {05}},\ \bibinfo {pages} {163}},\ \Eprint {https://arxiv.org/abs/1812.11192} {arXiv:1812.11192 [hep-ph]} \BibitemShut {NoStop}%
\bibitem [{\citenamefont {Co}\ \emph {et~al.}(2020{\natexlab{b}})\citenamefont {Co}, \citenamefont {Hall}, \citenamefont {Harigaya}, \citenamefont {Olive},\ and\ \citenamefont {Verner}}]{Co:2020dya}%
  \BibitemOpen
  \bibfield  {author} {\bibinfo {author} {\bibfnamefont {R.~T.}\ \bibnamefont {Co}}, \bibinfo {author} {\bibfnamefont {L.~J.}\ \bibnamefont {Hall}}, \bibinfo {author} {\bibfnamefont {K.}~\bibnamefont {Harigaya}}, \bibinfo {author} {\bibfnamefont {K.~A.}\ \bibnamefont {Olive}},\ and\ \bibinfo {author} {\bibfnamefont {S.}~\bibnamefont {Verner}},\ }\bibfield  {title} {\bibinfo {title} {{Axion Kinetic Misalignment and Parametric Resonance from Inflation}},\ }\href {https://doi.org/10.1088/1475-7516/2020/08/036} {\bibfield  {journal} {\bibinfo  {journal} {JCAP}\ }\textbf {\bibinfo {volume} {08}},\ \bibinfo {pages} {036}},\ \Eprint {https://arxiv.org/abs/2004.00629} {arXiv:2004.00629 [hep-ph]} \BibitemShut {NoStop}%
\bibitem [{\citenamefont {Cembranos}\ \emph {et~al.}(2020)\citenamefont {Cembranos}, \citenamefont {Garay},\ and\ \citenamefont {S\'anchez~Vel\'azquez}}]{Cembranos:2019qlm}%
  \BibitemOpen
  \bibfield  {author} {\bibinfo {author} {\bibfnamefont {J.~A.~R.}\ \bibnamefont {Cembranos}}, \bibinfo {author} {\bibfnamefont {L.~J.}\ \bibnamefont {Garay}},\ and\ \bibinfo {author} {\bibfnamefont {J.~M.}\ \bibnamefont {S\'anchez~Vel\'azquez}},\ }\bibfield  {title} {\bibinfo {title} {{Gravitational production of scalar dark matter}},\ }\href {https://doi.org/10.1007/JHEP06(2020)084} {\bibfield  {journal} {\bibinfo  {journal} {JHEP}\ }\textbf {\bibinfo {volume} {06}},\ \bibinfo {pages} {084}},\ \Eprint {https://arxiv.org/abs/1910.13937} {arXiv:1910.13937 [hep-ph]} \BibitemShut {NoStop}%
\bibitem [{\citenamefont {Boehm}\ \emph {et~al.}(2004)\citenamefont {Boehm}, \citenamefont {Ensslin},\ and\ \citenamefont {Silk}}]{Boehm:2002yz}%
  \BibitemOpen
  \bibfield  {author} {\bibinfo {author} {\bibfnamefont {C.}~\bibnamefont {Boehm}}, \bibinfo {author} {\bibfnamefont {T.~A.}\ \bibnamefont {Ensslin}},\ and\ \bibinfo {author} {\bibfnamefont {J.}~\bibnamefont {Silk}},\ }\bibfield  {title} {\bibinfo {title} {{Can Annihilating dark matter be lighter than a few GeVs?}},\ }\href {https://doi.org/10.1088/0954-3899/30/3/004} {\bibfield  {journal} {\bibinfo  {journal} {J. Phys. G}\ }\textbf {\bibinfo {volume} {30}},\ \bibinfo {pages} {279} (\bibinfo {year} {2004})},\ \Eprint {https://arxiv.org/abs/astro-ph/0208458} {arXiv:astro-ph/0208458} \BibitemShut {NoStop}%
\bibitem [{\citenamefont {Serpico}\ and\ \citenamefont {Raffelt}(2004)}]{Serpico:2004nm}%
  \BibitemOpen
  \bibfield  {author} {\bibinfo {author} {\bibfnamefont {P.~D.}\ \bibnamefont {Serpico}}\ and\ \bibinfo {author} {\bibfnamefont {G.~G.}\ \bibnamefont {Raffelt}},\ }\bibfield  {title} {\bibinfo {title} {{MeV-mass dark matter and primordial nucleosynthesis}},\ }\href {https://doi.org/10.1103/PhysRevD.70.043526} {\bibfield  {journal} {\bibinfo  {journal} {Phys. Rev. D}\ }\textbf {\bibinfo {volume} {70}},\ \bibinfo {pages} {043526} (\bibinfo {year} {2004})},\ \Eprint {https://arxiv.org/abs/astro-ph/0403417} {arXiv:astro-ph/0403417} \BibitemShut {NoStop}%
\bibitem [{\citenamefont {Nollett}\ and\ \citenamefont {Steigman}(2014)}]{Nollett:2013pwa}%
  \BibitemOpen
  \bibfield  {author} {\bibinfo {author} {\bibfnamefont {K.~M.}\ \bibnamefont {Nollett}}\ and\ \bibinfo {author} {\bibfnamefont {G.}~\bibnamefont {Steigman}},\ }\bibfield  {title} {\bibinfo {title} {{BBN And The CMB Constrain Light, Electromagnetically Coupled WIMPs}},\ }\href {https://doi.org/10.1103/PhysRevD.89.083508} {\bibfield  {journal} {\bibinfo  {journal} {Phys. Rev. D}\ }\textbf {\bibinfo {volume} {89}},\ \bibinfo {pages} {083508} (\bibinfo {year} {2014})},\ \Eprint {https://arxiv.org/abs/1312.5725} {arXiv:1312.5725 [astro-ph.CO]} \BibitemShut {NoStop}%
\bibitem [{\citenamefont {Steigman}\ and\ \citenamefont {Nollett}(2014)}]{Steigman:2014pfa}%
  \BibitemOpen
  \bibfield  {author} {\bibinfo {author} {\bibfnamefont {G.}~\bibnamefont {Steigman}}\ and\ \bibinfo {author} {\bibfnamefont {K.~M.}\ \bibnamefont {Nollett}},\ }\bibfield  {title} {\bibinfo {title} {{Light WIMPs, Equivalent Neutrinos, BBN, and the CMB}},\ }\href@noop {} {\bibfield  {journal} {\bibinfo  {journal} {Mem. Soc. Ast. It.}\ }\textbf {\bibinfo {volume} {85}},\ \bibinfo {pages} {175} (\bibinfo {year} {2014})},\ \Eprint {https://arxiv.org/abs/1401.5488} {arXiv:1401.5488 [astro-ph.CO]} \BibitemShut {NoStop}%
\bibitem [{\citenamefont {Nollett}\ and\ \citenamefont {Steigman}(2015)}]{Nollett:2014lwa}%
  \BibitemOpen
  \bibfield  {author} {\bibinfo {author} {\bibfnamefont {K.~M.}\ \bibnamefont {Nollett}}\ and\ \bibinfo {author} {\bibfnamefont {G.}~\bibnamefont {Steigman}},\ }\bibfield  {title} {\bibinfo {title} {{BBN And The CMB Constrain Neutrino Coupled Light WIMPs}},\ }\href {https://doi.org/10.1103/PhysRevD.91.083505} {\bibfield  {journal} {\bibinfo  {journal} {Phys. Rev. D}\ }\textbf {\bibinfo {volume} {91}},\ \bibinfo {pages} {083505} (\bibinfo {year} {2015})},\ \Eprint {https://arxiv.org/abs/1411.6005} {arXiv:1411.6005 [astro-ph.CO]} \BibitemShut {NoStop}%
\bibitem [{\citenamefont {Escudero}(2019)}]{Escudero:2018mvt}%
  \BibitemOpen
  \bibfield  {author} {\bibinfo {author} {\bibfnamefont {M.}~\bibnamefont {Escudero}},\ }\bibfield  {title} {\bibinfo {title} {{Neutrino decoupling beyond the Standard Model: CMB constraints on the Dark Matter mass with a fast and precise $N_{\rm eff}$ evaluation}},\ }\href {https://doi.org/10.1088/1475-7516/2019/02/007} {\bibfield  {journal} {\bibinfo  {journal} {JCAP}\ }\textbf {\bibinfo {volume} {02}},\ \bibinfo {pages} {007}},\ \Eprint {https://arxiv.org/abs/1812.05605} {arXiv:1812.05605 [hep-ph]} \BibitemShut {NoStop}%
\bibitem [{\citenamefont {Sabti}\ \emph {et~al.}(2020)\citenamefont {Sabti}, \citenamefont {Alvey}, \citenamefont {Escudero}, \citenamefont {Fairbairn},\ and\ \citenamefont {Blas}}]{Sabti:2019mhn}%
  \BibitemOpen
  \bibfield  {author} {\bibinfo {author} {\bibfnamefont {N.}~\bibnamefont {Sabti}}, \bibinfo {author} {\bibfnamefont {J.}~\bibnamefont {Alvey}}, \bibinfo {author} {\bibfnamefont {M.}~\bibnamefont {Escudero}}, \bibinfo {author} {\bibfnamefont {M.}~\bibnamefont {Fairbairn}},\ and\ \bibinfo {author} {\bibfnamefont {D.}~\bibnamefont {Blas}},\ }\bibfield  {title} {\bibinfo {title} {{Refined Bounds on MeV-scale Thermal Dark Sectors from BBN and the CMB}},\ }\href {https://doi.org/10.1088/1475-7516/2020/01/004} {\bibfield  {journal} {\bibinfo  {journal} {JCAP}\ }\textbf {\bibinfo {volume} {01}},\ \bibinfo {pages} {004}},\ \Eprint {https://arxiv.org/abs/1910.01649} {arXiv:1910.01649 [hep-ph]} \BibitemShut {NoStop}%
\bibitem [{\citenamefont {Giovanetti}\ \emph {et~al.}(2022)\citenamefont {Giovanetti}, \citenamefont {Lisanti}, \citenamefont {Liu},\ and\ \citenamefont {Ruderman}}]{Giovanetti:2021izc}%
  \BibitemOpen
  \bibfield  {author} {\bibinfo {author} {\bibfnamefont {C.}~\bibnamefont {Giovanetti}}, \bibinfo {author} {\bibfnamefont {M.}~\bibnamefont {Lisanti}}, \bibinfo {author} {\bibfnamefont {H.}~\bibnamefont {Liu}},\ and\ \bibinfo {author} {\bibfnamefont {J.~T.}\ \bibnamefont {Ruderman}},\ }\bibfield  {title} {\bibinfo {title} {{Joint Cosmic Microwave Background and Big Bang Nucleosynthesis Constraints on Light Dark Sectors with Dark Radiation}},\ }\href {https://doi.org/10.1103/PhysRevLett.129.021302} {\bibfield  {journal} {\bibinfo  {journal} {Phys. Rev. Lett.}\ }\textbf {\bibinfo {volume} {129}},\ \bibinfo {pages} {021302} (\bibinfo {year} {2022})},\ \Eprint {https://arxiv.org/abs/2109.03246} {arXiv:2109.03246 [hep-ph]} \BibitemShut {NoStop}%
\bibitem [{\citenamefont {An}\ \emph {et~al.}(2022)\citenamefont {An}, \citenamefont {Gluscevic}, \citenamefont {Calabrese},\ and\ \citenamefont {Hill}}]{An:2022sva}%
  \BibitemOpen
  \bibfield  {author} {\bibinfo {author} {\bibfnamefont {R.}~\bibnamefont {An}}, \bibinfo {author} {\bibfnamefont {V.}~\bibnamefont {Gluscevic}}, \bibinfo {author} {\bibfnamefont {E.}~\bibnamefont {Calabrese}},\ and\ \bibinfo {author} {\bibfnamefont {J.~C.}\ \bibnamefont {Hill}},\ }\bibfield  {title} {\bibinfo {title} {{What does cosmology tell us about the mass of thermal-relic dark matter?}},\ }\href {https://doi.org/10.1088/1475-7516/2022/07/002} {\bibfield  {journal} {\bibinfo  {journal} {JCAP}\ }\textbf {\bibinfo {volume} {07}}\bibfield  {number} {\bibinfo  {number} { (07)},\ \bibinfo {pages} {002}},\ }\Eprint {https://arxiv.org/abs/2202.03515} {arXiv:2202.03515 [astro-ph.CO]} \BibitemShut {NoStop}%
\bibitem [{\citenamefont {An}\ \emph {et~al.}(2024)\citenamefont {An}, \citenamefont {Boddy},\ and\ \citenamefont {Gluscevic}}]{An:2024nsw}%
  \BibitemOpen
  \bibfield  {author} {\bibinfo {author} {\bibfnamefont {R.}~\bibnamefont {An}}, \bibinfo {author} {\bibfnamefont {K.~K.}\ \bibnamefont {Boddy}},\ and\ \bibinfo {author} {\bibfnamefont {V.}~\bibnamefont {Gluscevic}},\ }\bibfield  {title} {\bibinfo {title} {{Interacting light thermal-relic dark matter: Self-consistent cosmological bounds}},\ }\href {https://doi.org/10.1103/PhysRevD.109.123522} {\bibfield  {journal} {\bibinfo  {journal} {Phys. Rev. D}\ }\textbf {\bibinfo {volume} {109}},\ \bibinfo {pages} {123522} (\bibinfo {year} {2024})},\ \Eprint {https://arxiv.org/abs/2402.14223} {arXiv:2402.14223 [astro-ph.CO]} \BibitemShut {NoStop}%
\bibitem [{\citenamefont {Mitridate}\ \emph {et~al.}(2021)\citenamefont {Mitridate}, \citenamefont {Trickle}, \citenamefont {Zhang},\ and\ \citenamefont {Zurek}}]{Mitridate:2021ctr}%
  \BibitemOpen
  \bibfield  {author} {\bibinfo {author} {\bibfnamefont {A.}~\bibnamefont {Mitridate}}, \bibinfo {author} {\bibfnamefont {T.}~\bibnamefont {Trickle}}, \bibinfo {author} {\bibfnamefont {Z.}~\bibnamefont {Zhang}},\ and\ \bibinfo {author} {\bibfnamefont {K.~M.}\ \bibnamefont {Zurek}},\ }\bibfield  {title} {\bibinfo {title} {{Dark matter absorption via electronic excitations}},\ }\href {https://doi.org/10.1007/JHEP09(2021)123} {\bibfield  {journal} {\bibinfo  {journal} {JHEP}\ }\textbf {\bibinfo {volume} {09}},\ \bibinfo {pages} {123}},\ \Eprint {https://arxiv.org/abs/2106.12586} {arXiv:2106.12586 [hep-ph]} \BibitemShut {NoStop}%
\bibitem [{\citenamefont {Knapen}\ \emph {et~al.}(2017)\citenamefont {Knapen}, \citenamefont {Lin},\ and\ \citenamefont {Zurek}}]{Knapen:2017xzo}%
  \BibitemOpen
  \bibfield  {author} {\bibinfo {author} {\bibfnamefont {S.}~\bibnamefont {Knapen}}, \bibinfo {author} {\bibfnamefont {T.}~\bibnamefont {Lin}},\ and\ \bibinfo {author} {\bibfnamefont {K.~M.}\ \bibnamefont {Zurek}},\ }\bibfield  {title} {\bibinfo {title} {{Light Dark Matter: Models and Constraints}},\ }\href {https://doi.org/10.1103/PhysRevD.96.115021} {\bibfield  {journal} {\bibinfo  {journal} {Phys. Rev. D}\ }\textbf {\bibinfo {volume} {96}},\ \bibinfo {pages} {115021} (\bibinfo {year} {2017})},\ \Eprint {https://arxiv.org/abs/1709.07882} {arXiv:1709.07882 [hep-ph]} \BibitemShut {NoStop}%
\bibitem [{\citenamefont {Essig}\ \emph {et~al.}(2013)\citenamefont {Essig}, \citenamefont {Kuflik}, \citenamefont {McDermott}, \citenamefont {Volansky},\ and\ \citenamefont {Zurek}}]{Essig:2013goa}%
  \BibitemOpen
  \bibfield  {author} {\bibinfo {author} {\bibfnamefont {R.}~\bibnamefont {Essig}}, \bibinfo {author} {\bibfnamefont {E.}~\bibnamefont {Kuflik}}, \bibinfo {author} {\bibfnamefont {S.~D.}\ \bibnamefont {McDermott}}, \bibinfo {author} {\bibfnamefont {T.}~\bibnamefont {Volansky}},\ and\ \bibinfo {author} {\bibfnamefont {K.~M.}\ \bibnamefont {Zurek}},\ }\bibfield  {title} {\bibinfo {title} {{Constraining Light Dark Matter with Diffuse X-Ray and Gamma-Ray Observations}},\ }\href {https://doi.org/10.1007/JHEP11(2013)193} {\bibfield  {journal} {\bibinfo  {journal} {JHEP}\ }\textbf {\bibinfo {volume} {11}},\ \bibinfo {pages} {193}},\ \Eprint {https://arxiv.org/abs/1309.4091} {arXiv:1309.4091 [hep-ph]} \BibitemShut {NoStop}%
\bibitem [{\citenamefont {Adams}\ \emph {et~al.}(1998)\citenamefont {Adams}, \citenamefont {Sarkar},\ and\ \citenamefont {Sciama}}]{Adams:1998nr}%
  \BibitemOpen
  \bibfield  {author} {\bibinfo {author} {\bibfnamefont {J.~A.}\ \bibnamefont {Adams}}, \bibinfo {author} {\bibfnamefont {S.}~\bibnamefont {Sarkar}},\ and\ \bibinfo {author} {\bibfnamefont {D.~W.}\ \bibnamefont {Sciama}},\ }\bibfield  {title} {\bibinfo {title} {{CMB anisotropy in the decaying neutrino cosmology}},\ }\href {https://doi.org/10.1046/j.1365-8711.1998.02017.x} {\bibfield  {journal} {\bibinfo  {journal} {Mon. Not. Roy. Astron. Soc.}\ }\textbf {\bibinfo {volume} {301}},\ \bibinfo {pages} {210} (\bibinfo {year} {1998})},\ \Eprint {https://arxiv.org/abs/astro-ph/9805108} {arXiv:astro-ph/9805108} \BibitemShut {NoStop}%
\bibitem [{\citenamefont {Chen}\ and\ \citenamefont {Kamionkowski}(2004)}]{Chen:2003gz}%
  \BibitemOpen
  \bibfield  {author} {\bibinfo {author} {\bibfnamefont {X.-L.}\ \bibnamefont {Chen}}\ and\ \bibinfo {author} {\bibfnamefont {M.}~\bibnamefont {Kamionkowski}},\ }\bibfield  {title} {\bibinfo {title} {{Particle decays during the cosmic dark ages}},\ }\href {https://doi.org/10.1103/PhysRevD.70.043502} {\bibfield  {journal} {\bibinfo  {journal} {Phys. Rev. D}\ }\textbf {\bibinfo {volume} {70}},\ \bibinfo {pages} {043502} (\bibinfo {year} {2004})},\ \Eprint {https://arxiv.org/abs/astro-ph/0310473} {arXiv:astro-ph/0310473} \BibitemShut {NoStop}%
\bibitem [{\citenamefont {{Finkbeiner}}\ \emph {et~al.}(2012)\citenamefont {{Finkbeiner}}, \citenamefont {{Galli}}, \citenamefont {{Lin}},\ and\ \citenamefont {{Slatyer}}}]{2012PhRvD..85d3522F}%
  \BibitemOpen
  \bibfield  {author} {\bibinfo {author} {\bibfnamefont {D.~P.}\ \bibnamefont {{Finkbeiner}}}, \bibinfo {author} {\bibfnamefont {S.}~\bibnamefont {{Galli}}}, \bibinfo {author} {\bibfnamefont {T.}~\bibnamefont {{Lin}}},\ and\ \bibinfo {author} {\bibfnamefont {T.~R.}\ \bibnamefont {{Slatyer}}},\ }\bibfield  {title} {\bibinfo {title} {{Searching for dark matter in the CMB: A compact parametrization of energy injection from new physics}},\ }\href {https://doi.org/10.1103/PhysRevD.85.043522} {\bibfield  {journal} {\bibinfo  {journal} {\prd}\ }\textbf {\bibinfo {volume} {85}},\ \bibinfo {eid} {043522} (\bibinfo {year} {2012})},\ \Eprint {https://arxiv.org/abs/1109.6322} {arXiv:1109.6322 [astro-ph.CO]} \BibitemShut {NoStop}%
\bibitem [{\citenamefont {Slatyer}(2013)}]{Slatyer:2012yq}%
  \BibitemOpen
  \bibfield  {author} {\bibinfo {author} {\bibfnamefont {T.~R.}\ \bibnamefont {Slatyer}},\ }\bibfield  {title} {\bibinfo {title} {{Energy Injection And Absorption In The Cosmic Dark Ages}},\ }\href {https://doi.org/10.1103/PhysRevD.87.123513} {\bibfield  {journal} {\bibinfo  {journal} {Phys. Rev. D}\ }\textbf {\bibinfo {volume} {87}},\ \bibinfo {pages} {123513} (\bibinfo {year} {2013})},\ \Eprint {https://arxiv.org/abs/1211.0283} {arXiv:1211.0283 [astro-ph.CO]} \BibitemShut {NoStop}%
\bibitem [{\citenamefont {Poulin}\ \emph {et~al.}(2016)\citenamefont {Poulin}, \citenamefont {Serpico},\ and\ \citenamefont {Lesgourgues}}]{Poulin:2016nat}%
  \BibitemOpen
  \bibfield  {author} {\bibinfo {author} {\bibfnamefont {V.}~\bibnamefont {Poulin}}, \bibinfo {author} {\bibfnamefont {P.~D.}\ \bibnamefont {Serpico}},\ and\ \bibinfo {author} {\bibfnamefont {J.}~\bibnamefont {Lesgourgues}},\ }\bibfield  {title} {\bibinfo {title} {{A fresh look at linear cosmological constraints on a decaying dark matter component}},\ }\href {https://doi.org/10.1088/1475-7516/2016/08/036} {\bibfield  {journal} {\bibinfo  {journal} {JCAP}\ }\textbf {\bibinfo {volume} {08}},\ \bibinfo {pages} {036}},\ \Eprint {https://arxiv.org/abs/1606.02073} {arXiv:1606.02073 [astro-ph.CO]} \BibitemShut {NoStop}%
\bibitem [{\citenamefont {Slatyer}\ and\ \citenamefont {Wu}(2017)}]{Slatyer:2016qyl}%
  \BibitemOpen
  \bibfield  {author} {\bibinfo {author} {\bibfnamefont {T.~R.}\ \bibnamefont {Slatyer}}\ and\ \bibinfo {author} {\bibfnamefont {C.-L.}\ \bibnamefont {Wu}},\ }\bibfield  {title} {\bibinfo {title} {{General Constraints on Dark Matter Decay from the Cosmic Microwave Background}},\ }\href {https://doi.org/10.1103/PhysRevD.95.023010} {\bibfield  {journal} {\bibinfo  {journal} {Phys. Rev. D}\ }\textbf {\bibinfo {volume} {95}},\ \bibinfo {pages} {023010} (\bibinfo {year} {2017})},\ \Eprint {https://arxiv.org/abs/1610.06933} {arXiv:1610.06933 [astro-ph.CO]} \BibitemShut {NoStop}%
\bibitem [{\citenamefont {Liu}\ \emph {et~al.}(2016)\citenamefont {Liu}, \citenamefont {Slatyer},\ and\ \citenamefont {Zavala}}]{Liu:2016cnk}%
  \BibitemOpen
  \bibfield  {author} {\bibinfo {author} {\bibfnamefont {H.}~\bibnamefont {Liu}}, \bibinfo {author} {\bibfnamefont {T.~R.}\ \bibnamefont {Slatyer}},\ and\ \bibinfo {author} {\bibfnamefont {J.}~\bibnamefont {Zavala}},\ }\bibfield  {title} {\bibinfo {title} {{Contributions to cosmic reionization from dark matter annihilation and decay}},\ }\href {https://doi.org/10.1103/PhysRevD.94.063507} {\bibfield  {journal} {\bibinfo  {journal} {Phys. Rev. D}\ }\textbf {\bibinfo {volume} {94}},\ \bibinfo {pages} {063507} (\bibinfo {year} {2016})},\ \Eprint {https://arxiv.org/abs/1604.02457} {arXiv:1604.02457 [astro-ph.CO]} \BibitemShut {NoStop}%
\bibitem [{\citenamefont {Poulin}\ \emph {et~al.}(2017)\citenamefont {Poulin}, \citenamefont {Lesgourgues},\ and\ \citenamefont {Serpico}}]{Poulin:2016anj}%
  \BibitemOpen
  \bibfield  {author} {\bibinfo {author} {\bibfnamefont {V.}~\bibnamefont {Poulin}}, \bibinfo {author} {\bibfnamefont {J.}~\bibnamefont {Lesgourgues}},\ and\ \bibinfo {author} {\bibfnamefont {P.~D.}\ \bibnamefont {Serpico}},\ }\bibfield  {title} {\bibinfo {title} {{Cosmological constraints on exotic injection of electromagnetic energy}},\ }\href {https://doi.org/10.1088/1475-7516/2017/03/043} {\bibfield  {journal} {\bibinfo  {journal} {JCAP}\ }\textbf {\bibinfo {volume} {03}},\ \bibinfo {pages} {043}},\ \Eprint {https://arxiv.org/abs/1610.10051} {arXiv:1610.10051 [astro-ph.CO]} \BibitemShut {NoStop}%
\bibitem [{\citenamefont {Cang}\ \emph {et~al.}(2020)\citenamefont {Cang}, \citenamefont {Gao},\ and\ \citenamefont {Ma}}]{Cang:2020exa}%
  \BibitemOpen
  \bibfield  {author} {\bibinfo {author} {\bibfnamefont {J.}~\bibnamefont {Cang}}, \bibinfo {author} {\bibfnamefont {Y.}~\bibnamefont {Gao}},\ and\ \bibinfo {author} {\bibfnamefont {Y.-Z.}\ \bibnamefont {Ma}},\ }\bibfield  {title} {\bibinfo {title} {{Probing dark matter with future CMB measurements}},\ }\href {https://doi.org/10.1103/PhysRevD.102.103005} {\bibfield  {journal} {\bibinfo  {journal} {Phys. Rev. D}\ }\textbf {\bibinfo {volume} {102}},\ \bibinfo {pages} {103005} (\bibinfo {year} {2020})},\ \Eprint {https://arxiv.org/abs/2002.03380} {arXiv:2002.03380 [astro-ph.CO]} \BibitemShut {NoStop}%
\bibitem [{\citenamefont {Liu}\ \emph {et~al.}(2023{\natexlab{a}})\citenamefont {Liu}, \citenamefont {Qin}, \citenamefont {Ridgway},\ and\ \citenamefont {Slatyer}}]{Liu:2023nct}%
  \BibitemOpen
  \bibfield  {author} {\bibinfo {author} {\bibfnamefont {H.}~\bibnamefont {Liu}}, \bibinfo {author} {\bibfnamefont {W.}~\bibnamefont {Qin}}, \bibinfo {author} {\bibfnamefont {G.~W.}\ \bibnamefont {Ridgway}},\ and\ \bibinfo {author} {\bibfnamefont {T.~R.}\ \bibnamefont {Slatyer}},\ }\bibfield  {title} {\bibinfo {title} {{Exotic energy injection in the early Universe. II. CMB spectral distortions and constraints on light dark matter}},\ }\href {https://doi.org/10.1103/PhysRevD.108.043531} {\bibfield  {journal} {\bibinfo  {journal} {Phys. Rev. D}\ }\textbf {\bibinfo {volume} {108}},\ \bibinfo {pages} {043531} (\bibinfo {year} {2023}{\natexlab{a}})},\ \Eprint {https://arxiv.org/abs/2303.07370} {arXiv:2303.07370 [astro-ph.CO]} \BibitemShut {NoStop}%
\bibitem [{\citenamefont {Capozzi}\ \emph {et~al.}(2023)\citenamefont {Capozzi}, \citenamefont {Ferreira}, \citenamefont {Lopez-Honorez},\ and\ \citenamefont {Mena}}]{Capozzi:2023xie}%
  \BibitemOpen
  \bibfield  {author} {\bibinfo {author} {\bibfnamefont {F.}~\bibnamefont {Capozzi}}, \bibinfo {author} {\bibfnamefont {R.~Z.}\ \bibnamefont {Ferreira}}, \bibinfo {author} {\bibfnamefont {L.}~\bibnamefont {Lopez-Honorez}},\ and\ \bibinfo {author} {\bibfnamefont {O.}~\bibnamefont {Mena}},\ }\bibfield  {title} {\bibinfo {title} {{CMB and Lyman-\ensuremath{\alpha} constraints on dark matter decays to photons}},\ }\href {https://doi.org/10.1088/1475-7516/2023/06/060} {\bibfield  {journal} {\bibinfo  {journal} {JCAP}\ }\textbf {\bibinfo {volume} {06}},\ \bibinfo {pages} {060}},\ \Eprint {https://arxiv.org/abs/2303.07426} {arXiv:2303.07426 [astro-ph.CO]} \BibitemShut {NoStop}%
\bibitem [{\citenamefont {{Xu}}\ \emph {et~al.}(2024)\citenamefont {{Xu}}, \citenamefont {{Qin}},\ and\ \citenamefont {{Slatyer}}}]{2024arXiv240813305X}%
  \BibitemOpen
  \bibfield  {author} {\bibinfo {author} {\bibfnamefont {C.}~\bibnamefont {{Xu}}}, \bibinfo {author} {\bibfnamefont {W.}~\bibnamefont {{Qin}}},\ and\ \bibinfo {author} {\bibfnamefont {T.~R.}\ \bibnamefont {{Slatyer}}},\ }\bibfield  {title} {\bibinfo {title} {{CMB limits on decaying dark matter beyond the ionization threshold}},\ }\href {https://doi.org/10.48550/arXiv.2408.13305} {\bibfield  {journal} {\bibinfo  {journal} {arXiv e-prints}\ ,\ \bibinfo {eid} {arXiv:2408.13305}} (\bibinfo {year} {2024})},\ \Eprint {https://arxiv.org/abs/2408.13305} {arXiv:2408.13305 [astro-ph.CO]} \BibitemShut {NoStop}%
\bibitem [{\citenamefont {Liu}\ \emph {et~al.}(2020)\citenamefont {Liu}, \citenamefont {Ridgway},\ and\ \citenamefont {Slatyer}}]{Liu:2019bbm}%
  \BibitemOpen
  \bibfield  {author} {\bibinfo {author} {\bibfnamefont {H.}~\bibnamefont {Liu}}, \bibinfo {author} {\bibfnamefont {G.~W.}\ \bibnamefont {Ridgway}},\ and\ \bibinfo {author} {\bibfnamefont {T.~R.}\ \bibnamefont {Slatyer}},\ }\bibfield  {title} {\bibinfo {title} {{Code package for calculating modified cosmic ionization and thermal histories with dark matter and other exotic energy injections}},\ }\href {https://doi.org/10.1103/PhysRevD.101.023530} {\bibfield  {journal} {\bibinfo  {journal} {Phys. Rev. D}\ }\textbf {\bibinfo {volume} {101}},\ \bibinfo {pages} {023530} (\bibinfo {year} {2020})},\ \Eprint {https://arxiv.org/abs/1904.09296} {arXiv:1904.09296 [astro-ph.CO]} \BibitemShut {NoStop}%
\bibitem [{\citenamefont {Liu}\ \emph {et~al.}(2023{\natexlab{b}})\citenamefont {Liu}, \citenamefont {Qin}, \citenamefont {Ridgway},\ and\ \citenamefont {Slatyer}}]{Liu:2023fgu}%
  \BibitemOpen
  \bibfield  {author} {\bibinfo {author} {\bibfnamefont {H.}~\bibnamefont {Liu}}, \bibinfo {author} {\bibfnamefont {W.}~\bibnamefont {Qin}}, \bibinfo {author} {\bibfnamefont {G.~W.}\ \bibnamefont {Ridgway}},\ and\ \bibinfo {author} {\bibfnamefont {T.~R.}\ \bibnamefont {Slatyer}},\ }\bibfield  {title} {\bibinfo {title} {{Exotic energy injection in the early Universe. I. A novel treatment for low-energy electrons and photons}},\ }\href {https://doi.org/10.1103/PhysRevD.108.043530} {\bibfield  {journal} {\bibinfo  {journal} {Phys. Rev. D}\ }\textbf {\bibinfo {volume} {108}},\ \bibinfo {pages} {043530} (\bibinfo {year} {2023}{\natexlab{b}})},\ \Eprint {https://arxiv.org/abs/2303.07366} {arXiv:2303.07366 [astro-ph.CO]} \BibitemShut {NoStop}%
\bibitem [{\citenamefont {St\"ocker}\ \emph {et~al.}(2018)\citenamefont {St\"ocker}, \citenamefont {Kr\"amer}, \citenamefont {Lesgourgues},\ and\ \citenamefont {Poulin}}]{Stocker:2018avm}%
  \BibitemOpen
  \bibfield  {author} {\bibinfo {author} {\bibfnamefont {P.}~\bibnamefont {St\"ocker}}, \bibinfo {author} {\bibfnamefont {M.}~\bibnamefont {Kr\"amer}}, \bibinfo {author} {\bibfnamefont {J.}~\bibnamefont {Lesgourgues}},\ and\ \bibinfo {author} {\bibfnamefont {V.}~\bibnamefont {Poulin}},\ }\bibfield  {title} {\bibinfo {title} {{Exotic energy injection with ExoCLASS: Application to the Higgs portal model and evaporating black holes}},\ }\href {https://doi.org/10.1088/1475-7516/2018/03/018} {\bibfield  {journal} {\bibinfo  {journal} {JCAP}\ }\textbf {\bibinfo {volume} {03}},\ \bibinfo {pages} {018}},\ \Eprint {https://arxiv.org/abs/1801.01871} {arXiv:1801.01871 [astro-ph.CO]} \BibitemShut {NoStop}%
\bibitem [{\citenamefont {Blas}\ \emph {et~al.}(2011)\citenamefont {Blas}, \citenamefont {Lesgourgues},\ and\ \citenamefont {Tram}}]{Blas:2011rf}%
  \BibitemOpen
  \bibfield  {author} {\bibinfo {author} {\bibfnamefont {D.}~\bibnamefont {Blas}}, \bibinfo {author} {\bibfnamefont {J.}~\bibnamefont {Lesgourgues}},\ and\ \bibinfo {author} {\bibfnamefont {T.}~\bibnamefont {Tram}},\ }\bibfield  {title} {\bibinfo {title} {{The Cosmic Linear Anisotropy Solving System~(CLASS)~II: Approximation Schemes}},\ }\href {https://doi.org/10.1088/1475-7516/2011/07/034} {\bibfield  {journal} {\bibinfo  {journal} {JCAP}\ }\textbf {\bibinfo {volume} {07}},\ \bibinfo {pages} {034}},\ \Eprint {https://arxiv.org/abs/1104.2933} {arXiv:1104.2933 [astro-ph.CO]} \BibitemShut {NoStop}%
\bibitem [{\citenamefont {Akrami}\ \emph {et~al.}(2020)\citenamefont {Akrami} \emph {et~al.}}]{Planck:2020olo}%
  \BibitemOpen
  \bibfield  {author} {\bibinfo {author} {\bibfnamefont {Y.}~\bibnamefont {Akrami}} \emph {et~al.} (\bibinfo {collaboration} {Planck}),\ }\bibfield  {title} {\bibinfo {title} {{$Planck$ intermediate results. LVII. Joint Planck LFI and HFI data processing}},\ }\href {https://doi.org/10.1051/0004-6361/202038073} {\bibfield  {journal} {\bibinfo  {journal} {Astron. Astrophys.}\ }\textbf {\bibinfo {volume} {643}},\ \bibinfo {pages} {A42} (\bibinfo {year} {2020})},\ \Eprint {https://arxiv.org/abs/2007.04997} {arXiv:2007.04997 [astro-ph.CO]} \BibitemShut {NoStop}%
\bibitem [{\citenamefont {Quevillon}\ and\ \citenamefont {Smith}(2019)}]{Quevillon:2019zrd}%
  \BibitemOpen
  \bibfield  {author} {\bibinfo {author} {\bibfnamefont {J.}~\bibnamefont {Quevillon}}\ and\ \bibinfo {author} {\bibfnamefont {C.}~\bibnamefont {Smith}},\ }\bibfield  {title} {\bibinfo {title} {{Axions are blind to anomalies}},\ }\href {https://doi.org/10.1140/epjc/s10052-019-7304-4} {\bibfield  {journal} {\bibinfo  {journal} {Eur. Phys. J. C}\ }\textbf {\bibinfo {volume} {79}},\ \bibinfo {pages} {822} (\bibinfo {year} {2019})},\ \Eprint {https://arxiv.org/abs/1903.12559} {arXiv:1903.12559 [hep-ph]} \BibitemShut {NoStop}%
\bibitem [{\citenamefont {Ferreira}\ \emph {et~al.}(2022{\natexlab{a}})\citenamefont {Ferreira}, \citenamefont {Marsh},\ and\ \citenamefont {M\"uller}}]{Ferreira:2022xlw}%
  \BibitemOpen
  \bibfield  {author} {\bibinfo {author} {\bibfnamefont {R.~Z.}\ \bibnamefont {Ferreira}}, \bibinfo {author} {\bibfnamefont {M.~C.~D.}\ \bibnamefont {Marsh}},\ and\ \bibinfo {author} {\bibfnamefont {E.}~\bibnamefont {M\"uller}},\ }\bibfield  {title} {\bibinfo {title} {{Strong supernovae bounds on ALPs from quantum loops}},\ }\href {https://doi.org/10.1088/1475-7516/2022/11/057} {\bibfield  {journal} {\bibinfo  {journal} {JCAP}\ }\textbf {\bibinfo {volume} {11}},\ \bibinfo {pages} {057}},\ \Eprint {https://arxiv.org/abs/2205.07896} {arXiv:2205.07896 [hep-ph]} \BibitemShut {NoStop}%
\bibitem [{\citenamefont {Green}\ \emph {et~al.}(2022)\citenamefont {Green}, \citenamefont {Guo},\ and\ \citenamefont {Wallisch}}]{Green:2021hjh}%
  \BibitemOpen
  \bibfield  {author} {\bibinfo {author} {\bibfnamefont {D.}~\bibnamefont {Green}}, \bibinfo {author} {\bibfnamefont {Y.}~\bibnamefont {Guo}},\ and\ \bibinfo {author} {\bibfnamefont {B.}~\bibnamefont {Wallisch}},\ }\bibfield  {title} {\bibinfo {title} {{Cosmological implications of axion-matter couplings}},\ }\href {https://doi.org/10.1088/1475-7516/2022/02/019} {\bibfield  {journal} {\bibinfo  {journal} {JCAP}\ }\textbf {\bibinfo {volume} {02}}\bibfield  {number} {\bibinfo  {number} { (02)},\ \bibinfo {pages} {019}},\ }\Eprint {https://arxiv.org/abs/2109.12088} {arXiv:2109.12088 [astro-ph.CO]} \BibitemShut {NoStop}%
\bibitem [{\citenamefont {Tristram}\ \emph {et~al.}(2024)\citenamefont {Tristram} \emph {et~al.}}]{Tristram:2023haj}%
  \BibitemOpen
  \bibfield  {author} {\bibinfo {author} {\bibfnamefont {M.}~\bibnamefont {Tristram}} \emph {et~al.},\ }\bibfield  {title} {\bibinfo {title} {{Cosmological parameters derived from the final Planck data release (PR4)}},\ }\href {https://doi.org/10.1051/0004-6361/202348015} {\bibfield  {journal} {\bibinfo  {journal} {Astron. Astrophys.}\ }\textbf {\bibinfo {volume} {682}},\ \bibinfo {pages} {A37} (\bibinfo {year} {2024})},\ \Eprint {https://arxiv.org/abs/2309.10034} {arXiv:2309.10034 [astro-ph.CO]} \BibitemShut {NoStop}%
\bibitem [{\citenamefont {{N. Aghanim \textit{et al.} (Planck Collaboration)}}(2020)}]{Planck:2019nip}%
  \BibitemOpen
  \bibfield  {author} {\bibinfo {author} {\bibnamefont {{N. Aghanim \textit{et al.} (Planck Collaboration)}}},\ }\bibfield  {title} {\bibinfo {title} {{Planck 2018 Results. V. CMB Power Spectra and Likelihoods}},\ }\href {https://doi.org/10.1051/0004-6361/201936386} {\bibfield  {journal} {\bibinfo  {journal} {Astron. Astrophys.}\ }\textbf {\bibinfo {volume} {641}},\ \bibinfo {pages} {A5} (\bibinfo {year} {2020})},\ \Eprint {https://arxiv.org/abs/1907.12875} {arXiv:1907.12875 [astro-ph.CO]} \BibitemShut {NoStop}%
\bibitem [{\citenamefont {Carron}\ \emph {et~al.}(2022)\citenamefont {Carron}, \citenamefont {Mirmelstein},\ and\ \citenamefont {Lewis}}]{Carron:2022eyg}%
  \BibitemOpen
  \bibfield  {author} {\bibinfo {author} {\bibfnamefont {J.}~\bibnamefont {Carron}}, \bibinfo {author} {\bibfnamefont {M.}~\bibnamefont {Mirmelstein}},\ and\ \bibinfo {author} {\bibfnamefont {A.}~\bibnamefont {Lewis}},\ }\bibfield  {title} {\bibinfo {title} {{CMB lensing from Planck PR4~maps}},\ }\href {https://doi.org/10.1088/1475-7516/2022/09/039} {\bibfield  {journal} {\bibinfo  {journal} {JCAP}\ }\textbf {\bibinfo {volume} {09}},\ \bibinfo {pages} {039}},\ \Eprint {https://arxiv.org/abs/2206.07773} {arXiv:2206.07773 [astro-ph.CO]} \BibitemShut {NoStop}%
\bibitem [{\citenamefont {Li}\ \emph {et~al.}(2018)\citenamefont {Li}, \citenamefont {Gluscevic}, \citenamefont {Boddy},\ and\ \citenamefont {Madhavacheril}}]{Li:2018zdm}%
  \BibitemOpen
  \bibfield  {author} {\bibinfo {author} {\bibfnamefont {Z.}~\bibnamefont {Li}}, \bibinfo {author} {\bibfnamefont {V.}~\bibnamefont {Gluscevic}}, \bibinfo {author} {\bibfnamefont {K.~K.}\ \bibnamefont {Boddy}},\ and\ \bibinfo {author} {\bibfnamefont {M.~S.}\ \bibnamefont {Madhavacheril}},\ }\bibfield  {title} {\bibinfo {title} {{Disentangling Dark Physics with Cosmic Microwave Background Experiments}},\ }\href {https://doi.org/10.1103/PhysRevD.98.123524} {\bibfield  {journal} {\bibinfo  {journal} {Phys. Rev. D}\ }\textbf {\bibinfo {volume} {98}},\ \bibinfo {pages} {123524} (\bibinfo {year} {2018})},\ \Eprint {https://arxiv.org/abs/1806.10165} {arXiv:1806.10165 [astro-ph.CO]} \BibitemShut {NoStop}%
\bibitem [{\citenamefont {Audren}\ \emph {et~al.}(2013)\citenamefont {Audren}, \citenamefont {Lesgourgues}, \citenamefont {Benabed},\ and\ \citenamefont {Prunet}}]{Audren:2012wb}%
  \BibitemOpen
  \bibfield  {author} {\bibinfo {author} {\bibfnamefont {B.}~\bibnamefont {Audren}}, \bibinfo {author} {\bibfnamefont {J.}~\bibnamefont {Lesgourgues}}, \bibinfo {author} {\bibfnamefont {K.}~\bibnamefont {Benabed}},\ and\ \bibinfo {author} {\bibfnamefont {S.}~\bibnamefont {Prunet}},\ }\bibfield  {title} {\bibinfo {title} {{Conservative Constraints on Early Cosmology: An Illustration of the MontePython Cosmological Parameter Inference Code}},\ }\href {https://doi.org/10.1088/1475-7516/2013/02/001} {\bibfield  {journal} {\bibinfo  {journal} {JCAP}\ }\textbf {\bibinfo {volume} {02}},\ \bibinfo {pages} {001}},\ \Eprint {https://arxiv.org/abs/1210.7183} {arXiv:1210.7183 [astro-ph.CO]} \BibitemShut {NoStop}%
\bibitem [{\citenamefont {Brinckmann}\ and\ \citenamefont {Lesgourgues}(2019)}]{Brinckmann:2018cvx}%
  \BibitemOpen
  \bibfield  {author} {\bibinfo {author} {\bibfnamefont {T.}~\bibnamefont {Brinckmann}}\ and\ \bibinfo {author} {\bibfnamefont {J.}~\bibnamefont {Lesgourgues}},\ }\bibfield  {title} {\bibinfo {title} {{MontePython 3: Boosted MCMC Sampler and Other Features}},\ }\href {https://doi.org/10.1016/j.dark.2018.100260} {\bibfield  {journal} {\bibinfo  {journal} {Phys. Dark Univ.}\ }\textbf {\bibinfo {volume} {24}},\ \bibinfo {pages} {100260} (\bibinfo {year} {2019})},\ \Eprint {https://arxiv.org/abs/1804.07261} {arXiv:1804.07261 [astro-ph.CO]} \BibitemShut {NoStop}%
\bibitem [{\citenamefont {Gelman}\ and\ \citenamefont {Rubin}(1992)}]{Gelman:1992zz}%
  \BibitemOpen
  \bibfield  {author} {\bibinfo {author} {\bibfnamefont {A.}~\bibnamefont {Gelman}}\ and\ \bibinfo {author} {\bibfnamefont {D.}~\bibnamefont {Rubin}},\ }\bibfield  {title} {\bibinfo {title} {{Inference from Iterative Simulation using Multiple Sequences}},\ }\href {https://doi.org/10.1214/ss/1177011136} {\bibfield  {journal} {\bibinfo  {journal} {Statist. Sci.}\ }\textbf {\bibinfo {volume} {7}},\ \bibinfo {pages} {457} (\bibinfo {year} {1992})}\BibitemShut {NoStop}%
\bibitem [{\citenamefont {{Lewis}}(2019)}]{2019arXiv191013970L}%
  \BibitemOpen
  \bibfield  {author} {\bibinfo {author} {\bibfnamefont {A.}~\bibnamefont {{Lewis}}},\ }\bibfield  {title} {\bibinfo {title} {{GetDist: a Python package for analysing Monte Carlo samples}},\ }\href {https://doi.org/10.48550/arXiv.1910.13970} {\bibfield  {journal} {\bibinfo  {journal} {arXiv e-prints}\ ,\ \bibinfo {eid} {arXiv:1910.13970}} (\bibinfo {year} {2019})},\ \Eprint {https://arxiv.org/abs/1910.13970} {arXiv:1910.13970 [astro-ph.IM]} \BibitemShut {NoStop}%
\bibitem [{\citenamefont {Ziegler}(2019)}]{Ziegler:2019gjr}%
  \BibitemOpen
  \bibfield  {author} {\bibinfo {author} {\bibfnamefont {R.}~\bibnamefont {Ziegler}},\ }\bibfield  {title} {\bibinfo {title} {{Flavored Axions}},\ }\href {https://doi.org/10.22323/1.347.0035} {\bibfield  {journal} {\bibinfo  {journal} {PoS}\ }\textbf {\bibinfo {volume} {CORFU2018}},\ \bibinfo {pages} {035} (\bibinfo {year} {2019})},\ \Eprint {https://arxiv.org/abs/1905.01084} {arXiv:1905.01084 [hep-ph]} \BibitemShut {NoStop}%
\bibitem [{PAN(2023)}]{PANCI2023137919}%
  \BibitemOpen
  \bibfield  {title} {\bibinfo {title} {Axion dark matter from lepton flavor-violating decays},\ }\href {https://doi.org/https://doi.org/10.1016/j.physletb.2023.137919} {\bibfield  {journal} {\bibinfo  {journal} {Physics Letters B}\ }\textbf {\bibinfo {volume} {841}},\ \bibinfo {pages} {137919} (\bibinfo {year} {2023})}\BibitemShut {NoStop}%
\bibitem [{\citenamefont {Ziegler}(2024)}]{Ziegler:2023aoe}%
  \BibitemOpen
  \bibfield  {author} {\bibinfo {author} {\bibfnamefont {R.}~\bibnamefont {Ziegler}},\ }\bibfield  {title} {\bibinfo {title} {{Flavor Probes of Axion Dark Matter}},\ }\href {https://doi.org/10.22323/1.431.0086} {\bibfield  {journal} {\bibinfo  {journal} {PoS}\ }\textbf {\bibinfo {volume} {DISCRETE2022}},\ \bibinfo {pages} {086} (\bibinfo {year} {2024})},\ \Eprint {https://arxiv.org/abs/2303.13353} {arXiv:2303.13353 [hep-ph]} \BibitemShut {NoStop}%
\bibitem [{\citenamefont {Bauer}\ \emph {et~al.}(2022)\citenamefont {Bauer}, \citenamefont {Neubert}, \citenamefont {Renner}, \citenamefont {Schnubel},\ and\ \citenamefont {Thamm}}]{Bauer:2021mvw}%
  \BibitemOpen
  \bibfield  {author} {\bibinfo {author} {\bibfnamefont {M.}~\bibnamefont {Bauer}}, \bibinfo {author} {\bibfnamefont {M.}~\bibnamefont {Neubert}}, \bibinfo {author} {\bibfnamefont {S.}~\bibnamefont {Renner}}, \bibinfo {author} {\bibfnamefont {M.}~\bibnamefont {Schnubel}},\ and\ \bibinfo {author} {\bibfnamefont {A.}~\bibnamefont {Thamm}},\ }\bibfield  {title} {\bibinfo {title} {{Flavor probes of axion-like particles}},\ }\href {https://doi.org/10.1007/JHEP09(2022)056} {\bibfield  {journal} {\bibinfo  {journal} {JHEP}\ }\textbf {\bibinfo {volume} {09}},\ \bibinfo {pages} {056}},\ \Eprint {https://arxiv.org/abs/2110.10698} {arXiv:2110.10698 [hep-ph]} \BibitemShut {NoStop}%
\bibitem [{\citenamefont {{Li}}\ \emph {et~al.}(2025)\citenamefont {{Li}}, \citenamefont {{Liu}},\ and\ \citenamefont {{Song}}}]{2025arXiv250106294L}%
  \BibitemOpen
  \bibfield  {author} {\bibinfo {author} {\bibfnamefont {H.}~\bibnamefont {{Li}}}, \bibinfo {author} {\bibfnamefont {Z.}~\bibnamefont {{Liu}}},\ and\ \bibinfo {author} {\bibfnamefont {N.}~\bibnamefont {{Song}}},\ }\bibfield  {title} {\bibinfo {title} {{Probing axion and muon-philic new physics with muon beam dump}},\ }\href {https://doi.org/10.48550/arXiv.2501.06294} {\bibfield  {journal} {\bibinfo  {journal} {arXiv e-prints}\ ,\ \bibinfo {eid} {arXiv:2501.06294}} (\bibinfo {year} {2025})},\ \Eprint {https://arxiv.org/abs/2501.06294} {arXiv:2501.06294 [hep-ph]} \BibitemShut {NoStop}%
\bibitem [{\citenamefont {Bollig}\ \emph {et~al.}(2020)\citenamefont {Bollig}, \citenamefont {DeRocco}, \citenamefont {Graham},\ and\ \citenamefont {Janka}}]{Bollig:2020xdr}%
  \BibitemOpen
  \bibfield  {author} {\bibinfo {author} {\bibfnamefont {R.}~\bibnamefont {Bollig}}, \bibinfo {author} {\bibfnamefont {W.}~\bibnamefont {DeRocco}}, \bibinfo {author} {\bibfnamefont {P.~W.}\ \bibnamefont {Graham}},\ and\ \bibinfo {author} {\bibfnamefont {H.-T.}\ \bibnamefont {Janka}},\ }\bibfield  {title} {\bibinfo {title} {{Muons in Supernovae: Implications for the Axion-Muon Coupling}},\ }\href {https://doi.org/10.1103/PhysRevLett.125.051104} {\bibfield  {journal} {\bibinfo  {journal} {Phys. Rev. Lett.}\ }\textbf {\bibinfo {volume} {125}},\ \bibinfo {pages} {051104} (\bibinfo {year} {2020})},\ \bibinfo {note} {[Erratum: Phys.Rev.Lett. 126, 189901 (2021)]},\ \Eprint {https://arxiv.org/abs/2005.07141} {arXiv:2005.07141 [hep-ph]} \BibitemShut {NoStop}%
\bibitem [{\citenamefont {Croon}\ \emph {et~al.}(2021)\citenamefont {Croon}, \citenamefont {Elor}, \citenamefont {Leane},\ and\ \citenamefont {McDermott}}]{Croon:2020lrf}%
  \BibitemOpen
  \bibfield  {author} {\bibinfo {author} {\bibfnamefont {D.}~\bibnamefont {Croon}}, \bibinfo {author} {\bibfnamefont {G.}~\bibnamefont {Elor}}, \bibinfo {author} {\bibfnamefont {R.~K.}\ \bibnamefont {Leane}},\ and\ \bibinfo {author} {\bibfnamefont {S.~D.}\ \bibnamefont {McDermott}},\ }\bibfield  {title} {\bibinfo {title} {{Supernova Muons: New Constraints on $Z$' Bosons, Axions and ALPs}},\ }\href {https://doi.org/10.1007/JHEP01(2021)107} {\bibfield  {journal} {\bibinfo  {journal} {JHEP}\ }\textbf {\bibinfo {volume} {01}},\ \bibinfo {pages} {107}},\ \Eprint {https://arxiv.org/abs/2006.13942} {arXiv:2006.13942 [hep-ph]} \BibitemShut {NoStop}%
\bibitem [{\citenamefont {Caputo}\ \emph {et~al.}(2022)\citenamefont {Caputo}, \citenamefont {Raffelt},\ and\ \citenamefont {Vitagliano}}]{Caputo:2021rux}%
  \BibitemOpen
  \bibfield  {author} {\bibinfo {author} {\bibfnamefont {A.}~\bibnamefont {Caputo}}, \bibinfo {author} {\bibfnamefont {G.}~\bibnamefont {Raffelt}},\ and\ \bibinfo {author} {\bibfnamefont {E.}~\bibnamefont {Vitagliano}},\ }\bibfield  {title} {\bibinfo {title} {{Muonic boson limits: Supernova redux}},\ }\href {https://doi.org/10.1103/PhysRevD.105.035022} {\bibfield  {journal} {\bibinfo  {journal} {Phys. Rev. D}\ }\textbf {\bibinfo {volume} {105}},\ \bibinfo {pages} {035022} (\bibinfo {year} {2022})},\ \Eprint {https://arxiv.org/abs/2109.03244} {arXiv:2109.03244 [hep-ph]} \BibitemShut {NoStop}%
\bibitem [{\citenamefont {Hardy}\ and\ \citenamefont {Lasenby}(2017)}]{Hardy:2016kme}%
  \BibitemOpen
  \bibfield  {author} {\bibinfo {author} {\bibfnamefont {E.}~\bibnamefont {Hardy}}\ and\ \bibinfo {author} {\bibfnamefont {R.}~\bibnamefont {Lasenby}},\ }\bibfield  {title} {\bibinfo {title} {{Stellar cooling bounds on new light particles: plasma mixing effects}},\ }\href {https://doi.org/10.1007/JHEP02(2017)033} {\bibfield  {journal} {\bibinfo  {journal} {JHEP}\ }\textbf {\bibinfo {volume} {02}},\ \bibinfo {pages} {033}},\ \Eprint {https://arxiv.org/abs/1611.05852} {arXiv:1611.05852 [hep-ph]} \BibitemShut {NoStop}%
\bibitem [{\citenamefont {Ng}\ \emph {et~al.}(2019)\citenamefont {Ng}, \citenamefont {Roach}, \citenamefont {Perez}, \citenamefont {Beacom}, \citenamefont {Horiuchi}, \citenamefont {Krivonos},\ and\ \citenamefont {Wik}}]{Ng:2019gch}%
  \BibitemOpen
  \bibfield  {author} {\bibinfo {author} {\bibfnamefont {K.~C.~Y.}\ \bibnamefont {Ng}}, \bibinfo {author} {\bibfnamefont {B.~M.}\ \bibnamefont {Roach}}, \bibinfo {author} {\bibfnamefont {K.}~\bibnamefont {Perez}}, \bibinfo {author} {\bibfnamefont {J.~F.}\ \bibnamefont {Beacom}}, \bibinfo {author} {\bibfnamefont {S.}~\bibnamefont {Horiuchi}}, \bibinfo {author} {\bibfnamefont {R.}~\bibnamefont {Krivonos}},\ and\ \bibinfo {author} {\bibfnamefont {D.~R.}\ \bibnamefont {Wik}},\ }\bibfield  {title} {\bibinfo {title} {{New Constraints on Sterile Neutrino Dark Matter from $NuSTAR$ M31 Observations}},\ }\href {https://doi.org/10.1103/PhysRevD.99.083005} {\bibfield  {journal} {\bibinfo  {journal} {Phys. Rev. D}\ }\textbf {\bibinfo {volume} {99}},\ \bibinfo {pages} {083005} (\bibinfo {year} {2019})},\ \Eprint {https://arxiv.org/abs/1901.01262} {arXiv:1901.01262 [astro-ph.HE]} \BibitemShut {NoStop}%
\bibitem [{\citenamefont {Laha}\ \emph {et~al.}(2020)\citenamefont {Laha}, \citenamefont {Mu\~noz},\ and\ \citenamefont {Slatyer}}]{Laha:2020ivk}%
  \BibitemOpen
  \bibfield  {author} {\bibinfo {author} {\bibfnamefont {R.}~\bibnamefont {Laha}}, \bibinfo {author} {\bibfnamefont {J.~B.}\ \bibnamefont {Mu\~noz}},\ and\ \bibinfo {author} {\bibfnamefont {T.~R.}\ \bibnamefont {Slatyer}},\ }\bibfield  {title} {\bibinfo {title} {{INTEGRAL constraints on primordial black holes and particle dark matter}},\ }\href {https://doi.org/10.1103/PhysRevD.101.123514} {\bibfield  {journal} {\bibinfo  {journal} {Phys. Rev. D}\ }\textbf {\bibinfo {volume} {101}},\ \bibinfo {pages} {123514} (\bibinfo {year} {2020})},\ \Eprint {https://arxiv.org/abs/2004.00627} {arXiv:2004.00627 [astro-ph.CO]} \BibitemShut {NoStop}%
\bibitem [{\citenamefont {Foster}\ \emph {et~al.}(2021)\citenamefont {Foster}, \citenamefont {Kongsore}, \citenamefont {Dessert}, \citenamefont {Park}, \citenamefont {Rodd}, \citenamefont {Cranmer},\ and\ \citenamefont {Safdi}}]{Foster:2021ngm}%
  \BibitemOpen
  \bibfield  {author} {\bibinfo {author} {\bibfnamefont {J.~W.}\ \bibnamefont {Foster}}, \bibinfo {author} {\bibfnamefont {M.}~\bibnamefont {Kongsore}}, \bibinfo {author} {\bibfnamefont {C.}~\bibnamefont {Dessert}}, \bibinfo {author} {\bibfnamefont {Y.}~\bibnamefont {Park}}, \bibinfo {author} {\bibfnamefont {N.~L.}\ \bibnamefont {Rodd}}, \bibinfo {author} {\bibfnamefont {K.}~\bibnamefont {Cranmer}},\ and\ \bibinfo {author} {\bibfnamefont {B.~R.}\ \bibnamefont {Safdi}},\ }\bibfield  {title} {\bibinfo {title} {{Deep Search for Decaying Dark Matter with XMM-Newton Blank-Sky Observations}},\ }\href {https://doi.org/10.1103/PhysRevLett.127.051101} {\bibfield  {journal} {\bibinfo  {journal} {Phys. Rev. Lett.}\ }\textbf {\bibinfo {volume} {127}},\ \bibinfo {pages} {051101} (\bibinfo {year} {2021})},\ \Eprint {https://arxiv.org/abs/2102.02207} {arXiv:2102.02207 [astro-ph.CO]} \BibitemShut {NoStop}%
\bibitem [{\citenamefont {Ferreira}\ \emph {et~al.}(2022{\natexlab{b}})\citenamefont {Ferreira}, \citenamefont {Marsh},\ and\ \citenamefont {M\"uller}}]{Ferreira:2022egk}%
  \BibitemOpen
  \bibfield  {author} {\bibinfo {author} {\bibfnamefont {R.~Z.}\ \bibnamefont {Ferreira}}, \bibinfo {author} {\bibfnamefont {M.~C.~D.}\ \bibnamefont {Marsh}},\ and\ \bibinfo {author} {\bibfnamefont {E.}~\bibnamefont {M\"uller}},\ }\bibfield  {title} {\bibinfo {title} {{Do Direct Detection Experiments Constrain Axionlike Particles Coupled to Electrons?}},\ }\href {https://doi.org/10.1103/PhysRevLett.128.221302} {\bibfield  {journal} {\bibinfo  {journal} {Phys. Rev. Lett.}\ }\textbf {\bibinfo {volume} {128}},\ \bibinfo {pages} {221302} (\bibinfo {year} {2022}{\natexlab{b}})},\ \Eprint {https://arxiv.org/abs/2202.08858} {arXiv:2202.08858 [hep-ph]} \BibitemShut {NoStop}%
\bibitem [{\citenamefont {Bottaro}\ \emph {et~al.}(2023)\citenamefont {Bottaro}, \citenamefont {Caputo}, \citenamefont {Raffelt},\ and\ \citenamefont {Vitagliano}}]{Bottaro:2023gep}%
  \BibitemOpen
  \bibfield  {author} {\bibinfo {author} {\bibfnamefont {S.}~\bibnamefont {Bottaro}}, \bibinfo {author} {\bibfnamefont {A.}~\bibnamefont {Caputo}}, \bibinfo {author} {\bibfnamefont {G.}~\bibnamefont {Raffelt}},\ and\ \bibinfo {author} {\bibfnamefont {E.}~\bibnamefont {Vitagliano}},\ }\bibfield  {title} {\bibinfo {title} {{Stellar limits on scalars from electron-nucleus bremsstrahlung}},\ }\href {https://doi.org/10.1088/1475-7516/2023/07/071} {\bibfield  {journal} {\bibinfo  {journal} {JCAP}\ }\textbf {\bibinfo {volume} {07}},\ \bibinfo {pages} {071}},\ \Eprint {https://arxiv.org/abs/2303.00778} {arXiv:2303.00778 [hep-ph]} \BibitemShut {NoStop}%
\bibitem [{\citenamefont {Capozzi}\ and\ \citenamefont {Raffelt}(2020)}]{Capozzi:2020cbu}%
  \BibitemOpen
  \bibfield  {author} {\bibinfo {author} {\bibfnamefont {F.}~\bibnamefont {Capozzi}}\ and\ \bibinfo {author} {\bibfnamefont {G.}~\bibnamefont {Raffelt}},\ }\bibfield  {title} {\bibinfo {title} {{Axion and neutrino bounds improved with new calibrations of the tip of the red-giant branch using geometric distance determinations}},\ }\href {https://doi.org/10.1103/PhysRevD.102.083007} {\bibfield  {journal} {\bibinfo  {journal} {Phys. Rev. D}\ }\textbf {\bibinfo {volume} {102}},\ \bibinfo {pages} {083007} (\bibinfo {year} {2020})},\ \Eprint {https://arxiv.org/abs/2007.03694} {arXiv:2007.03694 [astro-ph.SR]} \BibitemShut {NoStop}%
\bibitem [{\citenamefont {Agnes}\ \emph {et~al.}(2023)\citenamefont {Agnes} \emph {et~al.}}]{DarkSide:2022knj}%
  \BibitemOpen
  \bibfield  {author} {\bibinfo {author} {\bibfnamefont {P.}~\bibnamefont {Agnes}} \emph {et~al.} (\bibinfo {collaboration} {DarkSide}),\ }\bibfield  {title} {\bibinfo {title} {{Search for Dark Matter Particle Interactions with Electron Final States with DarkSide-50}},\ }\href {https://doi.org/10.1103/PhysRevLett.130.101002} {\bibfield  {journal} {\bibinfo  {journal} {Phys. Rev. Lett.}\ }\textbf {\bibinfo {volume} {130}},\ \bibinfo {pages} {101002} (\bibinfo {year} {2023})},\ \Eprint {https://arxiv.org/abs/2207.11968} {arXiv:2207.11968 [hep-ex]} \BibitemShut {NoStop}%
\bibitem [{\citenamefont {Zeng}\ \emph {et~al.}(2024)\citenamefont {Zeng} \emph {et~al.}}]{PandaX:2024cic}%
  \BibitemOpen
  \bibfield  {author} {\bibinfo {author} {\bibfnamefont {X.}~\bibnamefont {Zeng}} \emph {et~al.} (\bibinfo {collaboration} {PandaX}),\ }\href@noop {} {\bibinfo {title} {{Exploring New Physics with PandaX-4T Low Energy Electronic Recoil Data}}} (\bibinfo {year} {2024}),\ \Eprint {https://arxiv.org/abs/2408.07641} {arXiv:2408.07641 [hep-ex]} \BibitemShut {NoStop}%
\bibitem [{\citenamefont {Li}\ \emph {et~al.}(2024)\citenamefont {Li} \emph {et~al.}}]{PandaX:2024kjp}%
  \BibitemOpen
  \bibfield  {author} {\bibinfo {author} {\bibfnamefont {T.}~\bibnamefont {Li}} \emph {et~al.} (\bibinfo {collaboration} {PandaX}),\ }\href@noop {} {\bibinfo {title} {{Searching for MeV-scale Axion-like Particles and Dark Photons with PandaX-4T}}} (\bibinfo {year} {2024}),\ \Eprint {https://arxiv.org/abs/2409.00773} {arXiv:2409.00773 [hep-ex]} \BibitemShut {NoStop}%
\bibitem [{\citenamefont {Aprile}\ \emph {et~al.}(2019)\citenamefont {Aprile} \emph {et~al.}}]{XENON:2019gfn}%
  \BibitemOpen
  \bibfield  {author} {\bibinfo {author} {\bibfnamefont {E.}~\bibnamefont {Aprile}} \emph {et~al.} (\bibinfo {collaboration} {XENON}),\ }\bibfield  {title} {\bibinfo {title} {{Light Dark Matter Search with Ionization Signals in XENON1T}},\ }\href {https://doi.org/10.1103/PhysRevLett.123.251801} {\bibfield  {journal} {\bibinfo  {journal} {Phys. Rev. Lett.}\ }\textbf {\bibinfo {volume} {123}},\ \bibinfo {pages} {251801} (\bibinfo {year} {2019})},\ \Eprint {https://arxiv.org/abs/1907.11485} {arXiv:1907.11485 [hep-ex]} \BibitemShut {NoStop}%
\bibitem [{\citenamefont {Aprile}\ \emph {et~al.}(2020)\citenamefont {Aprile} \emph {et~al.}}]{XENON:2020rca}%
  \BibitemOpen
  \bibfield  {author} {\bibinfo {author} {\bibfnamefont {E.}~\bibnamefont {Aprile}} \emph {et~al.} (\bibinfo {collaboration} {XENON}),\ }\bibfield  {title} {\bibinfo {title} {{Excess electronic recoil events in XENON1T}},\ }\href {https://doi.org/10.1103/PhysRevD.102.072004} {\bibfield  {journal} {\bibinfo  {journal} {Phys. Rev. D}\ }\textbf {\bibinfo {volume} {102}},\ \bibinfo {pages} {072004} (\bibinfo {year} {2020})},\ \Eprint {https://arxiv.org/abs/2006.09721} {arXiv:2006.09721 [hep-ex]} \BibitemShut {NoStop}%
\bibitem [{\citenamefont {Aprile}\ \emph {et~al.}(2022{\natexlab{a}})\citenamefont {Aprile} \emph {et~al.}}]{XENON:2021myl}%
  \BibitemOpen
  \bibfield  {author} {\bibinfo {author} {\bibfnamefont {E.}~\bibnamefont {Aprile}} \emph {et~al.} (\bibinfo {collaboration} {(XENON Collaboration)\textsection{}, XENON}),\ }\bibfield  {title} {\bibinfo {title} {{Emission of single and few electrons in XENON1T and limits on light dark matter}},\ }\href {https://doi.org/10.1103/PhysRevD.106.022001} {\bibfield  {journal} {\bibinfo  {journal} {Phys. Rev. D}\ }\textbf {\bibinfo {volume} {106}},\ \bibinfo {pages} {022001} (\bibinfo {year} {2022}{\natexlab{a}})},\ \Eprint {https://arxiv.org/abs/2112.12116} {arXiv:2112.12116 [hep-ex]} \BibitemShut {NoStop}%
\bibitem [{\citenamefont {Aprile}\ \emph {et~al.}(2022{\natexlab{b}})\citenamefont {Aprile} \emph {et~al.}}]{XENONCollaboration:2022kmb}%
  \BibitemOpen
  \bibfield  {author} {\bibinfo {author} {\bibfnamefont {E.}~\bibnamefont {Aprile}} \emph {et~al.} (\bibinfo {collaboration} {(XENON Collaboration)\textdagger{}\textdagger{}, XENON}),\ }\bibfield  {title} {\bibinfo {title} {{Search for New Physics in Electronic Recoil Data from XENONnT}},\ }\href {https://doi.org/10.1103/PhysRevLett.129.161805} {\bibfield  {journal} {\bibinfo  {journal} {Phys. Rev. Lett.}\ }\textbf {\bibinfo {volume} {129}},\ \bibinfo {pages} {161805} (\bibinfo {year} {2022}{\natexlab{b}})},\ \Eprint {https://arxiv.org/abs/2207.11330} {arXiv:2207.11330 [hep-ex]} \BibitemShut {NoStop}%
\bibitem [{\citenamefont {Aralis}\ \emph {et~al.}(2020)\citenamefont {Aralis} \emph {et~al.}}]{Aralis:2019nfa}%
  \BibitemOpen
  \bibfield  {author} {\bibinfo {author} {\bibfnamefont {T.}~\bibnamefont {Aralis}} \emph {et~al.} (\bibinfo {collaboration} {SuperCDMS}),\ }\bibfield  {title} {\bibinfo {title} {{Constraints on dark photons and axionlike particles from the SuperCDMS Soudan experiment}},\ }\href {https://doi.org/10.1103/PhysRevD.101.052008} {\bibfield  {journal} {\bibinfo  {journal} {Phys. Rev. D}\ }\textbf {\bibinfo {volume} {101}},\ \bibinfo {pages} {052008} (\bibinfo {year} {2020})},\ \bibinfo {note} {[Erratum: Phys.Rev.D 103, 039901 (2021)]},\ \Eprint {https://arxiv.org/abs/1911.11905} {arXiv:1911.11905 [hep-ex]} \BibitemShut {NoStop}%
\bibitem [{\citenamefont {Armengaud}\ \emph {et~al.}(2018)\citenamefont {Armengaud} \emph {et~al.}}]{EDELWEISS:2018tde}%
  \BibitemOpen
  \bibfield  {author} {\bibinfo {author} {\bibfnamefont {E.}~\bibnamefont {Armengaud}} \emph {et~al.} (\bibinfo {collaboration} {EDELWEISS}),\ }\bibfield  {title} {\bibinfo {title} {{Searches for electron interactions induced by new physics in the EDELWEISS-III Germanium bolometers}},\ }\href {https://doi.org/10.1103/PhysRevD.98.082004} {\bibfield  {journal} {\bibinfo  {journal} {Phys. Rev. D}\ }\textbf {\bibinfo {volume} {98}},\ \bibinfo {pages} {082004} (\bibinfo {year} {2018})},\ \Eprint {https://arxiv.org/abs/1808.02340} {arXiv:1808.02340 [hep-ex]} \BibitemShut {NoStop}%
\bibitem [{\citenamefont {Agostini}\ \emph {et~al.}(2020)\citenamefont {Agostini} \emph {et~al.}}]{GERDA:2020emj}%
  \BibitemOpen
  \bibfield  {author} {\bibinfo {author} {\bibfnamefont {M.}~\bibnamefont {Agostini}} \emph {et~al.} (\bibinfo {collaboration} {GERDA}),\ }\bibfield  {title} {\bibinfo {title} {{First Search for Bosonic Superweakly Interacting Massive Particles with Masses up to 1 MeV/$c^2$ with GERDA}},\ }\href {https://doi.org/10.1103/PhysRevLett.125.011801} {\bibfield  {journal} {\bibinfo  {journal} {Phys. Rev. Lett.}\ }\textbf {\bibinfo {volume} {125}},\ \bibinfo {pages} {011801} (\bibinfo {year} {2020})},\ \bibinfo {note} {[Erratum: Phys.Rev.Lett. 129, 089901 (2022)]},\ \Eprint {https://arxiv.org/abs/2005.14184} {arXiv:2005.14184 [hep-ex]} \BibitemShut {NoStop}%
\bibitem [{\citenamefont {P\'{e}rez}\ and\ \citenamefont {Granger}(2007)}]{Perez:2007ipy}%
  \BibitemOpen
  \bibfield  {author} {\bibinfo {author} {\bibfnamefont {F.}~\bibnamefont {P\'{e}rez}}\ and\ \bibinfo {author} {\bibfnamefont {B.}~\bibnamefont {Granger}},\ }\bibfield  {title} {\bibinfo {title} {{IPython: A System for Interactive Scientific Computing}},\ }\href {https://doi.org/10.1109/MCSE.2007.53} {\bibfield  {journal} {\bibinfo  {journal} {Comput. Sci. Eng.}\ }\textbf {\bibinfo {volume} {9}},\ \bibinfo {pages} {21} (\bibinfo {year} {2007})}\BibitemShut {NoStop}%
\bibitem [{\citenamefont {Hunter}(2007)}]{Hunter:2007mat}%
  \BibitemOpen
  \bibfield  {author} {\bibinfo {author} {\bibfnamefont {J.}~\bibnamefont {Hunter}},\ }\bibfield  {title} {\bibinfo {title} {{Matplotlib: A 2D Graphics Environment}},\ }\href {https://doi.org/10.1109/MCSE.2007.55} {\bibfield  {journal} {\bibinfo  {journal} {Comput. Sci. Eng.}\ }\textbf {\bibinfo {volume} {9}},\ \bibinfo {pages} {90} (\bibinfo {year} {2007})}\BibitemShut {NoStop}%
\bibitem [{\citenamefont {Harris}\ \emph {et~al.}(2020)\citenamefont {Harris} \emph {et~al.}}]{Harris:2020xlr}%
  \BibitemOpen
  \bibfield  {author} {\bibinfo {author} {\bibfnamefont {C.}~\bibnamefont {Harris}} \emph {et~al.},\ }\bibfield  {title} {\bibinfo {title} {{Array Programming with NumPy}},\ }\href {https://doi.org/10.1038/s41586-020-2649-2} {\bibfield  {journal} {\bibinfo  {journal} {Nature}\ }\textbf {\bibinfo {volume} {585}},\ \bibinfo {pages} {3572} (\bibinfo {year} {2020})},\ \Eprint {https://arxiv.org/abs/2006.10256} {arXiv:2006.10256 [cs.MS]} \BibitemShut {NoStop}%
\bibitem [{\citenamefont {Virtanen}\ \emph {et~al.}(2020)\citenamefont {Virtanen} \emph {et~al.}}]{Virtanen:2019joe}%
  \BibitemOpen
  \bibfield  {author} {\bibinfo {author} {\bibfnamefont {P.}~\bibnamefont {Virtanen}} \emph {et~al.},\ }\bibfield  {title} {\bibinfo {title} {{SciPy 1.0 -- Fundamental Algorithms for Scientific Computing in Python}},\ }\href {https://doi.org/10.1038/s41592-019-0686-2} {\bibfield  {journal} {\bibinfo  {journal} {Nat. Methods}\ }\textbf {\bibinfo {volume} {17}},\ \bibinfo {pages} {261} (\bibinfo {year} {2020})},\ \Eprint {https://arxiv.org/abs/1907.10121} {arXiv:1907.10121 [cs.MS]} \BibitemShut {NoStop}%
\bibitem [{\citenamefont {O'Hare}(2020)}]{AxionLimits}%
  \BibitemOpen
  \bibfield  {author} {\bibinfo {author} {\bibfnamefont {C.}~\bibnamefont {O'Hare}},\ }\href {https://doi.org/10.5281/zenodo.3932430} {\bibinfo {title} {cajohare/axionlimits: Axionlimits}},\ \bibinfo {howpublished} {\url{https://cajohare.github.io/AxionLimits/}} (\bibinfo {year} {2020})\BibitemShut {NoStop}%
\end{thebibliography}%
\clearpage

\onecolumngrid

\FloatBarrier


\begin{appendices}

\begin{figure}[ht!]
    \centering
    \includegraphics[width=\linewidth]{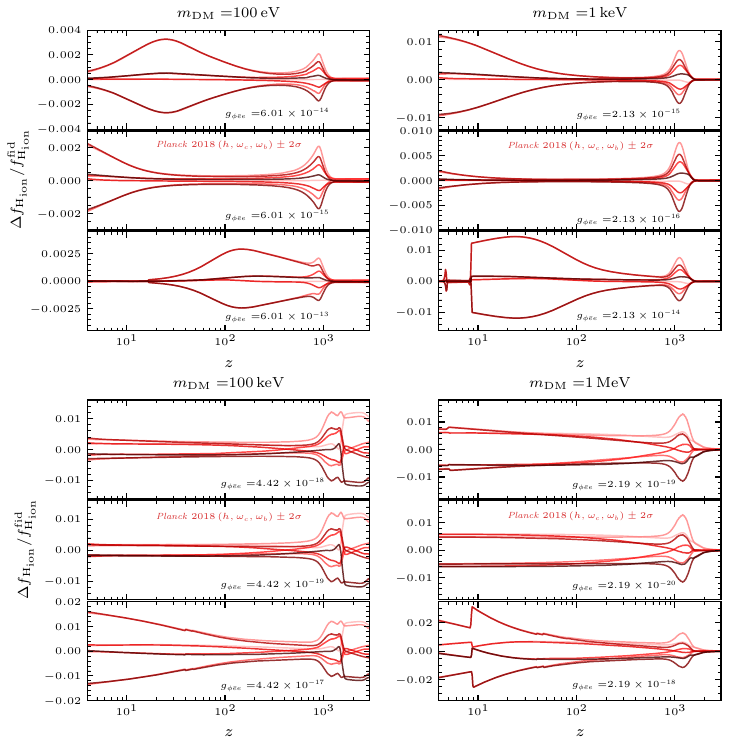}
    \caption{Same as in Fig.~\ref{fig:fs_cosmo_mDM10keV}, but for the four other DM masses used in our analysis, namely $m_{\mathrm{DM}}=100\,\mathrm{eV}$~({\it top left}), $m_{\mathrm{DM}}=1\,\mathrm{keV}$~({\it top right}), $m_{\mathrm{DM}}=100\,\mathrm{keV}$~({\it bottom left}), and $m_{\mathrm{DM}}=1\,\mathrm{MeV}$~({\it bottom right}). Each panel shows the residuals in the hydrogen ionization channel of the energy deposition function, $f_{\mathrm{H\,ion}}(z)$, for scalar DM-electron couplings $g_{\phi \bar{e}e}$ at the 95\%~C.L.\ exclusion limit from Ref.~\cite{2024arXiv240813305X} (\textit{top}), an order of magnitude smaller (\textit{middle}), and an order of magnitude larger (\textit{bottom}), all plotted against the \textit{Planck}~2018 best-fit cosmology. The eight curves in each panel, shown in various shades of \textcolor{tabred}{red}, represent different combinations of $(h,\,\omega_b,\,\omega_c)$ taken at their $\pm 95\%$~C.L.\ \textit{Planck}~2018 values. In all cases, cosmology-driven changes in $f_{\mathrm{H\,ion}}(z)$ remain roughly at the $\mathcal{O}(1)\%$ level near recombination for every mass considered, while at lower redshifts closer to reionization,  stronger couplings generally result in mildly larger residuals.}
    \label{fig:fs_cosmo_mDMs}
\end{figure}
\section{Cosmology Dependence of the energy deposition functions}
\label{app:fs_cosmo}

In this appendix, we present for completeness the cosmology-driven variations in the hydrogen ionization channel of the energy deposition function, $f_{\rm H_{ion}}(z)$, for the four additional DM masses analyzed in this work, namely $m_{\mathrm{DM}}$ of $100\,\mathrm{eV}$, $1\,\mathrm{keV}$, $100\,\mathrm{keV}$, and $1\,\mathrm{MeV}$. In Fig.~\ref{fig:fs_cosmo_mDMs} (analogous to Fig.~\ref{fig:fs_cosmo_mDM10keV}), we display, for each DM mass, the residuals in $f_{\rm H_{ion}}(z)$ computed under eight different cosmologies, with $(h,\,\omega_b,\,\omega_c)$ taken at their $\pm 95\%$~C.L.\ \textit{Planck}~2018 limits. For each mass, we consider three representative choices of the scalar DM–electron coupling, $g_{\phi \bar{e}e}$: a value that yields a DM lifetime at the 95\%~C.L.\ exclusion limit from Ref.~\cite{2024arXiv240813305X} and values that are one order of magnitude larger and smaller. While the precise residuals depend on both the DM mass and coupling (e.g., slightly larger deviations for DM masses $>1\,$keV and for stronger couplings at lower redshifts), we consistently find $f_{\rm H_{ion}}(z)$ changes remain at roughly the $\mathcal{O}(1)\%$ level near recombination. 
These small but non-zero departures motivate our cosmology-consistent analysis, $\textbf{PR3-}f^{\rm CC}_c$, although a comprehensive examination of their detailed origins lies beyond the scope of this work. In practice, the outcome of our $\textbf{PR3-}f^{\rm CC}_c$ study (see Sec.~\ref{sec:results}) demonstrates that these mild variations have no impact on the derived DM--electron coupling constraints and can thus be safely neglected in subsequent analyses of DM decay.

\begin{figure*}[tph!]
    \centering
    \includegraphics[width=.95\linewidth]{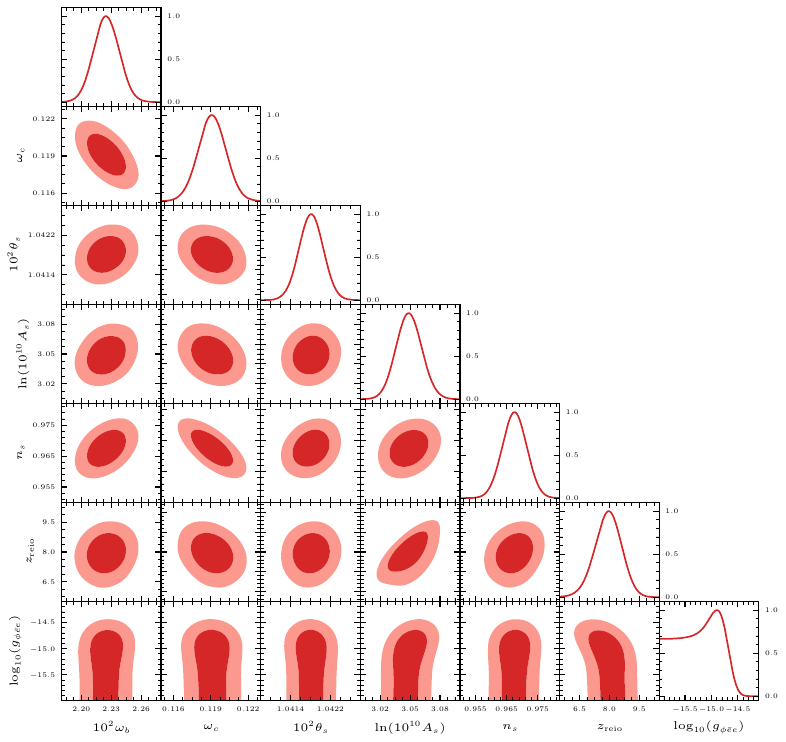}
    \caption{Marginalized 2D posterior distributions for the $\Lambda$CDM parameters and the dimensionless DM-electron coupling, $g_{\phi \bar e e}$, for a DM mass of $1\,$keV. Contours correspond to the 68\% and 95\% confidence levels, derived from a joint analysis of the \textit{Planck} PR4 likelihoods, incorporating temperature, polarization, and lensing anisotropies, as described in Sec.~\ref{sec:analysis}. The 1D marginalized posteriors for each parameter are displayed along the diagonal.}
    \label{fig:triangle_plot_mDM1keV_PR4}
\end{figure*}

\section{Posterior Distributions and Parameter Degeneracies}
\label{app:MCMC_analysis}
\begin{figure}
    \centering
    \includegraphics[width=.95\linewidth]{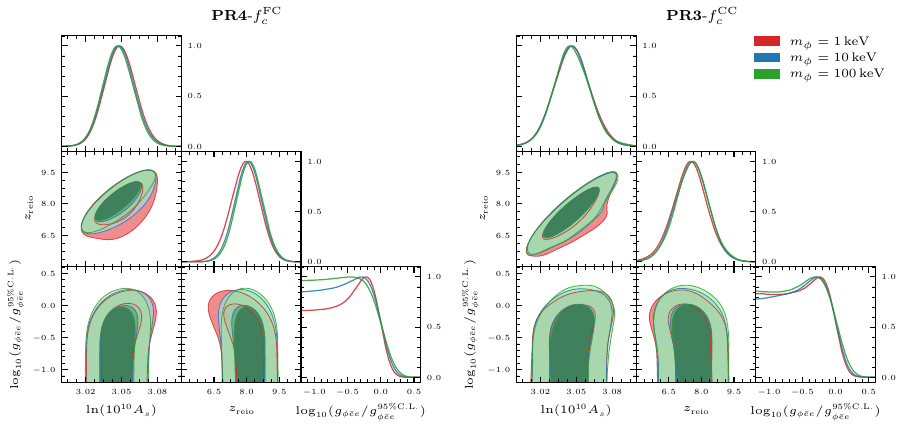}
    \caption{
    \textit{Left:} Marginalized 2D posterior distributions for $z_{\rm reio}$, $\ln(10^{10}A_s)$, and the DM-electron coupling $g_{\phi \bar e e}$ in the \textbf{PR4-$f^\mathrm{FC}_c$} analysis, where the energy deposition functions $f_c(z)$ are pre-computed under a fiducial cosmology. \textit{Right:} The equivalent triangle plot for the \textbf{PR3-$f^\mathrm{CC}_c$} analysis, where instead the $f_c(z)$ functions computed dynamically at each MCMC step. In both panels, we show the results for DM masses of $1\,\mathrm{keV}$ (\textcolor{tabred}{red}), $10\,\mathrm{keV}$ (\textcolor{tabblue}{blue}), and $100\,\mathrm{keV}$ (\textcolor{tabgreen}{green}), and rescale $g_{\phi \bar e e}$ by its respective 95\%\,C.L.\ upper limit, i.e.\ we plot $\log_{10}\bigl(g_{\phi \bar e e} / g_{\phi \bar e e}^{95\%\,\mathrm{C.L.}}\bigr)$. Contours correspond to the 68\% and 95\% confidence levels, derived from the joint analyses using the respective \textit{Planck} likelihoods (see Sec.~\ref{sec:analysis}), and 1D marginalized posteriors for each parameter appear along the diagonal.}
    \label{fig:PR4_degeneracies}
\end{figure}
In this appendix, we present representative examples of the reconstructed marginalized posterior probability distributions from our MCMC analyses and comment on degeneracies between parameters. 
We show in Fig.~\ref{fig:triangle_plot_mDM1keV_PR4} the triangle plot of the 2D posterior distributions for the $\Lambda$CDM parameters \{$10^2\omega_b$, $\omega_c$, $10^2\theta_s$, $\ln(10^{10}A_s)$, $n_s$ and $z_{\rm reio}$\} and the dimensionless DM-electron coupling $g_{\phi \bar e e}$, for a DM mass of $1\,\rm{keV}$. 
These results are derived from the \textit{Planck} PR4 likelihood analysis (\textbf{PR4-$f^\mathrm{FC}_c$}), described in Sec.~\ref{sec:analysis}, where we follow the standard treatment of energy injection, using pre-computed energy deposition functions~$f_c(z)$ under a fiducial cosmology. 

As expected, the constraints on the standard~$\Lambda$CDM parameters remain consistent with those from a $\Lambda$CDM-only analysis of the \textit{Planck} PR4 data~\cite{Tristram:2023haj}.
In addition, we note a mild degeneracy between $z_{\rm reio}$, $g_{\phi \bar e e}$, and $\ln(10^{10}A_s)$. 
The end of reionization is delayed in the presence of additional processes that contribute to ionizing gas, leading to a negative correlation between $g_{\phi \bar e e}$ and $z_{\rm reio}$. Although one might expect extra ionization to anticipate the end of reionization, partial ionization from DM decays at earlier times effectively reduces the astrophysical reionization required at high redshift. 
 In the MCMC fit, this translates into a lower $z_{\rm reio}$ to avoid overestimating the total optical depth, leading to a small negative correlation between $g_{\phi \bar e e}$ and $z_{\rm reio}$. Furthermore, the optical depth at reionization is closely tied to the overall amplitude of the CMB anisotropies, creating the known positive correlation between the amplitude of the primordial scalar power spectrum $A_s$ and $z_{\rm reio}$. 
As a result, a weak degeneracy between $g_{\phi \bar e e}$ and $\ln(10^{10}A_s)$ is induced through their mutual dependence on $z_{\rm reio}$. 

In Fig.~\ref{fig:PR4_degeneracies} (left panel), we further explore these degeneracies by comparing the joint posterior distributions of $z_{\rm reio}$, $g_{\phi \bar e e}$, and $\ln(10^{10}A_s)$ within the same \textbf{PR4-$f^{\rm FC}_c$} analysis for DM masses of $1\,$keV, $10\,$keV, and $100\,$keV.
We rescale DM-electron coupling for each mass by the respective 95\% C.L.\ upper limit [i.e., we show $\log_{10}(g_{\phi \bar e e}/g_{\phi \bar e e}^{95\%\,\rm{C.L.}})$], to more easily compare the 1D and 2D posterior distributions in a single plot.
Overall, this comparison reveals that the degeneracies among~$z_{\rm reio}$,~$g_{\phi \bar e e}$, and~$\ln(10^{10}A_s)$ gradually diminish as the DM mass increases, manifesting as flatter 1D posteriors for~$g_{\phi \bar e e}$ at values well below the 95\% C.L.\ upper limit for~$m_{\rm{DM}} \gtrsim 10\,\rm{keV}$.

In the right panel of Fig.~\ref{fig:PR4_degeneracies}, we show the equivalent triangle plot for the \textbf{PR3-$f^\mathrm{CC}_c$} analysis, for which we use the \textit{Planck} PR3 likelihood code and compute the~$f_c(z)$ functions dynamically at each MCMC step to reflect variations in cosmological parameters. Two main observations emerge from comparing this cosmology-consistent approach with the fixed-cosmology PR4 analysis (\textbf{PR4-$f^\mathrm{FC}_c$}) in the left panel. First, although the degeneracies among~$z_{\rm reio}$,~$g_{\phi \bar e e}$, and~$\ln(10^{10}A_s)$ remain qualitatively similar,  they are visibly weaker under the cosmology-consistent treatment, indicating that pre-computing $f_c(z)$ from a fiducial cosmology induces a mild extra degeneracy at low DM masses. Nevertheless, we find that this extra degeneracy does not substantially impact the final constraints, which remain consistent between the two approaches, as shown clearly in Fig.~\ref{fig:CMB_limits}. Second, the overall trend of diminishing degeneracies at larger DM masses persists in both PR3 and PR4 analyses, suggesting that the reduced parameter correlations for $m_{\rm DM}\gtrsim 10\,$keV may contribute to the small but noticeable improvement in the \textit{Planck} PR4–derived constraints.  Conversely, at lower DM masses, these degeneracies are not negligible, potentially limiting the impact of the enhanced polarization data in the updated PR4 likelihoods.

\end{appendices}

\end{document}